\documentclass{aa}

\usepackage{txfonts}
\usepackage{graphicx}
\usepackage{natbib}
\bibpunct{(}{)}{;}{a}{}{,}

\begin{document}

\title{The galaxy major merger fraction to $z \sim 1$}
\author{Carlos L\'opez-Sanjuan\inst{1} \and Marc Balcells\inst{1} \and Pablo 
G. P\'erez-Gonz\'alez\inst{2} \and Guillermo Barro\inst{2} \and C\'esar Enrique 
Garc\'{i}a-Dab\'o\inst{1,3} \and Jes\'us Gallego\inst{2} \and Jaime Zamorano\inst{2}}

\institute{Instituto de Astrof\'{\i}sica de Canarias, Calle V\'{\i}a L\'actea 
s/n, E-38205 La Laguna, Tenerife, Spain \and Departamento de Astrof\'{\i}sica y 
Ciencias de la Atm\'osfera, Facultad de C.C. F\'{\i}sicas, Universidad Complutense 
de Madrid, E-28040 Madrid, Spain \and European South Observatory, 
Karl-Schwarzschild-Strasse 2, D-85748 Garching, Germany} 

\date{Received 23 February 2009; Accepted 12 May 2009}

\abstract
{} {The importance of disc--disc major mergers in galaxy evolution remains uncertain. We study the major merger fraction in a 
SPITZER/IRAC-selected catalogue in the GOODS-S field up to $z \sim 1$ for 
luminosity- and mass-limited samples.} 
{We select disc--disc merger remnants on the basis of morphological asymmetries/distortions, and address three main sources of 
systematic errors: (i) we explicitly apply morphological K-corrections, (ii) we measure 
asymmetries in galaxies artificially redshifted to $z_{\rm d} = 1.0$ to deal 
with loss of morphological information with redshift, and (iii) we take into 
account the observational errors in $z$ and $A$, which tend to overestimate 
the merger fraction, though use of maximum likelihood techniques.} {We obtain 
morphological merger fractions ($f_{\rm m}^{\rm mph}$) below 0.06 up 
to $z \sim 1$. Parameterizing the merger fraction evolution with redshift as 
$f_{\rm m}^{\rm mph}(z) = f_{\rm m}^{\rm mph}(0) (1+z)^m$, we find that $m = 
1.8 \pm 0.5$ for $M_B \leq -20$ galaxies, while $m = 5.4 \pm 0.4$ for $M_{\star} 
\geq 10^{10}\ M_{\odot}$ galaxies. When we translate our merger fractions to 
merger rates ($\Re_{\rm m}^{\rm mph}$), their evolution, parameterized as $\Re_{\rm m}^{\rm mph}(z) = 
\Re_{\rm m}^{\rm mph}(0) (1+z)^n$, is quite similar in both cases: $n = 3.3 
\pm 0.8$ for $M_B \leq -20$ galaxies, and $n = 3.5 \pm 0.4$ for $M_{\star} 
\geq 10^{10}\ M_{\odot}$ galaxies.} {Our results imply that only $\sim8$\% 
of today's $M_{\star} \geq 10^{10}\ M_{\odot}$ galaxies have undergone a 
disc--disc major merger since $z \sim 1$. In addition, $\sim 21$\% of $M_{\star} \geq 10^{10}\ M_{\odot}$ galaxies at $z \sim 1$ have undergone one of these mergers since $z\sim1.5$. 
This suggests that disc--disc major mergers are not the dominant process in the 
evolution of $M_{\star} \geq 10^{10}\ M_{\odot}$ galaxies since $z\sim1$, with 
only 0.2 disc--disc major mergers per galaxy, but may be an important process 
at $z > 1$, with $\sim 1$ merger per galaxy at $1 < z < 3$.}


\maketitle

\section{INTRODUCTION}
The colour--magnitude diagram of local galaxies shows two distinct populations: 
the red sequence, consisting primarily of old, spheroid-dominated, quiescent galaxies, and 
the blue cloud, formed primarily by spiral and irregular star-forming galaxies \citep{strateva01, 
baldry04}. This bimodality has been traced at increasingly higher redshifts (\citealt{bell04}, 
up to $z \sim 1$; \citealt{arnouts07,cirasuolo07}, up to $z \sim 1.5$; 
\citealt{giallongo05, cassata08}, up to $z \sim 2$; \citealt{kriek08}, at 
$z \sim 2.3$). More massive galaxies were the first to populate the red sequence as a result of the so-called ''downsizing'' 
\citep{cowie96}: massive galaxies experienced most of their star formation 
at early times and are passive by $z \sim 1$, while many of the less massive galaxies 
have extended star formation histories \citep[see][and references therein]{bundy06,scarlata07ee,pgon08}.

These results pose a challenge to the popular hierarchical $\Lambda$-CDM models, 
in which one expects that the more massive dark matter halos are the final 
stage of successive minor halo mergers. However, the treatment of the baryonic 
component is still unclear. The latest models, which include radiative cooling, 
star formation, and AGN and supernova feedback, seem to reproduce the observational 
trends better \citep[see][and references therein]{bower06,delucia07,stewart09,hopkins09bulges}. 
Within this framework, the role of galaxy mergers in the build-up of the red sequence 
and their relative importance in the evolution of galaxy properties, i.e. colour,
 mass, or morphology, is an important open question.

The merger fraction, $f_{\rm m}$, defined as the ratio between the number of merger events in a
sample and the total number of sources in the same sample, is a useful observational quantity for answering that question. Many 
studies have determined the merger fraction and its evolution with redshift, 
usually parameterized as $f_{\rm m}(z) = f_{\rm m}(0) (1+z)^m$, using different
sample selections and methods, such as morphological criteria \citep[]{conselice03ff, 
conselice08, conselice09cos, lavery04, cassata05, lotz08ff, bridge07, kamp07, jogee09},
 kinematic close companions \citep[]{patton00, patton02, patton08, lin04, lin08, 
depropris05, depropris07, bluck09}, spatially close pairs \citep[]{lefevre00, bundy04,
 bundy09, bridge07, kar07, hsieh08}, or the correlation function \citep[]{bell06, 
masjedi06}. In these studies the value of the merger index $m$ at redshift $z 
\lesssim 1$ varies in the range $m =$ 0--4. $\Lambda$-CDM models predict $m 
\sim$ 2--3 \citep[]{kolatt99, governato99, gottlober01,fak08} for dark matter 
halos, while suggesting a weaker evolution, $m \sim$ 0--2, for the galaxy 
merger fraction \citep{berrier06,stewart08}.

To constrain the role of disc--disc major mergers in galaxy evolution, in this paper
 we study their redshift evolution up to $z \sim 1$ in a SPITZER/IRAC-selected 
catalogue of the GOODS-S area. We use morphological criteria, based on the fact that, 
just after a merger is complete, the galaxy image shows strong geometrical distortions, 
particularly asymmetric distortions \citep{conselice03}. Hence, high values in the 
automatic asymmetry index $A$ \citep{abraham96, conselice00} are assumed to identify 
disc--disc major merger systems. This methodology presents several systematic effects, such
 as signal-to-noise dependence \citep{conselice03,conselice05} or contamination by 
non-interacting galaxies with high asymmetry values \citep{jogee09,miller08}, which 
lead to biased merger fractions if not treated carefully. In a previous study of the 
Groth field, \citet[][L09 hereafter]{clsj09ffgs} demonstrated a robust procedure to determine
 morphological merger fractions ($f_{\rm m}^{\rm mph}$) using galaxy asymmetries. In that study 
they avoid the loss of information with redshift by artificially moving all sources 
to a common redshift, while the experimental error bias, which tends to overestimate the 
merger fraction up to 50\%, was addressed through use of a maximum likelihood method developed in 
\citet[][LGB08 hereafter]{clsj08ml}. L09 find that the merger rate 
decreases with stellar mass at $z = 0.6$, and that 20--35\% of present-day $M_B \leq 
-20$ galaxies have undergone a disc--disc major merger since $z \sim 1$.

This paper is organized as follows: in Sect.~\ref{data} we summarize the GOODS-S data 
set that we use in our study, and in Sect.~\ref{asy} we develop the asymmetry index 
calculations and study their variation with redshift. Then, in Sect.~\ref{metodo} we 
use the methodology to obtain the morphological merger fraction by taking into account 
the observational errors. In Sect.~\ref{results} we summarize the obtained merger 
fractions and their evolution with $z$, while in Sect.~\ref{discussion} we compare
 our results with other authors. Finally, in Sect.~\ref{conclusion} we present our
 conclusions. We use $H_0 = 70\ {\rm km\ s^{-1}\ Mpc^{-1}}$, $\Omega_{M} = 0.3$, 
and $\Omega_{\Lambda} = 0.7$ throughout. All magnitudes are Vega unless noted otherwise.

\section{DATA}\label{data}
\subsection{The GOODS-S SPITZER/IRAC-selected catalogue}
This work is based on the analysis of the structural parameters of the galaxies 
catalogued in the GOODS-South field by the Spitzer Legacy Team \citep[see][]{giavalisco04}. 
We used the Version 1.0 catalogues\footnote{http://archive.stsci.edu/prepds/goods/} 
and reduced mosaics in the $F435W$ ($B_{435}$), $F606W$ ($V_{606}$), $F775W$ ($i_{775}$), 
and $F850LP$ ($z_{850}$) HST/ACS bands. These catalogues were cross-correlated 
using a $1.5^{\prime\prime}$ search radius with the GOODS-S IRAC-selected sample
 in the Rainbow Cosmological Database published in \citet[see also \citealt{pgon05} 
and Barro et al. 2009, in prep.]{pgon08}, which provided us with spectral energy
 distributions (SEDs) in the UV-to-MIR range, well-calibrated and reliable 
photometric redshifts, stellar masses, star formation rates and rest-frame absolute magnitudes.

We refer the reader to the above-mentioned papers for a more detailed description of 
the data included in the SEDs and the analysis procedure. Here, we summarize briefly 
the main characteristics of the data set. We measured consistent aperture photometry 
in several UV, optical, NIR and MIR bands with the method described in \citet{pgon08}. 
UV-to-MIR SEDs were built for all IRAC sources in the GOODS-S region down to a 75\% 
completeness magnitude $[3.6]$$=$23.5~mag (AB). These SEDs were fitted to stellar 
population and dust emission models to obtain an estimate of the photometric 
redshift ($z_{\rm phot}$), the stellar mass ($M_{\star}$), and the rest-frame absolute 
B-band magnitude ($M_B$).

The median accuracy of the photometric redshifts at 
$z < 1.5$ is $|z_{\rm spec} - z_{\rm phot}|/(1+z_{\rm spec}) = 0.04$, 
with a fraction $<$5\% of catastrophic outliers \citep[][Fig.~B2]{pgon08}.
Rest-frame absolute B-band magnitudes were estimated for each source by convolving 
the templates fitting the SED with the transmission curve of a typical Bessel $B$ filter, 
taking into account the redshift of each source. This procedure provided us 
with accurately interpolated $B$-band magnitudes including a robustly estimated 
$k$-correction. Stellar masses were estimated using exponential star formation 
PEGASE01 models with a \citet{salpeter55} IMF and various ages, metallicities 
and dust contents included. The typical uncertainties in the stellar masses 
are a factor of $\sim$2 (typical of any stellar 
population study; see, e.g., \citealt{papovich06}, \citealt{fontana06}).

Finally, our methodology requires the errors in $z_{\rm phot}$ to be Gaussian 
(Sect.~\ref{metodo}, LGB08, L09), while $z_{\rm phot}$ confidence intervals 
given by $\chi^2$ methods do not correlate with the differences between $z_{\rm spec}$'s 
and $z_{\rm phot}$'s \citep{oyaizu08}. Because of this, and following L09, we use 
$\sigma_{z_{\rm phot}} = \sigma_{\delta_z} (1+z_{\rm phot})$ as the 
$z_{\rm phot}$ error, where $\sigma_{\delta_z}$ is the standard deviation in 
the distribution of the variable $\delta_z \equiv (z_{\rm phot} - z_{\rm spec}) 
/ ({1 + z_{\rm phot}})$, which is well described by a Gaussian with mean $\mu_{\delta_z} 
\sim 0$ and standard deviation $\sigma_{\delta_z}$. We found that $\sigma_{\delta_z}$ 
increases with redshift, and we took $\sigma_{\delta_z} = 0.043$ for $z \leq 0.9$ 
sources and $\sigma_{\delta_z} = 0.05$ for $z > 0.9$ sources. This procedure assigns 
the same error to sources with equal $z_{\rm phot}$, but it is statistically 
representative of our sample and ensures the best Gaussian approximation of 
$z_{\rm phot}$ errors in the merger fraction determination (Sect.~\ref{metodo}).

\begin{figure}[t]
\resizebox{\hsize}{!}{\includegraphics{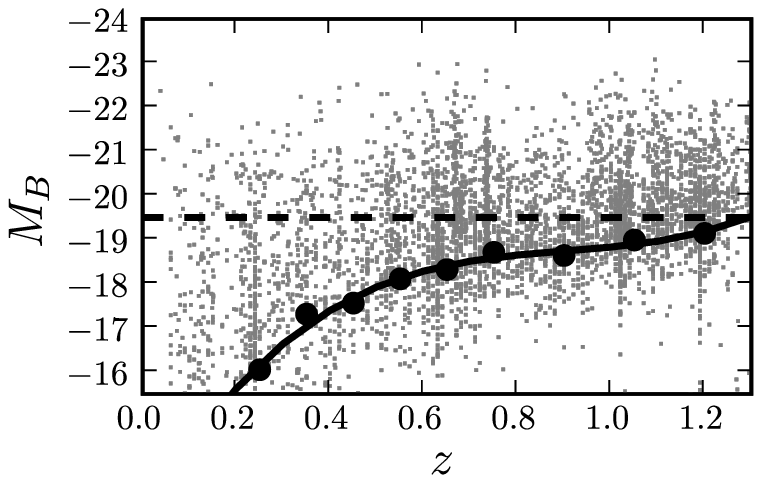}}
\resizebox{\hsize}{!}{\includegraphics{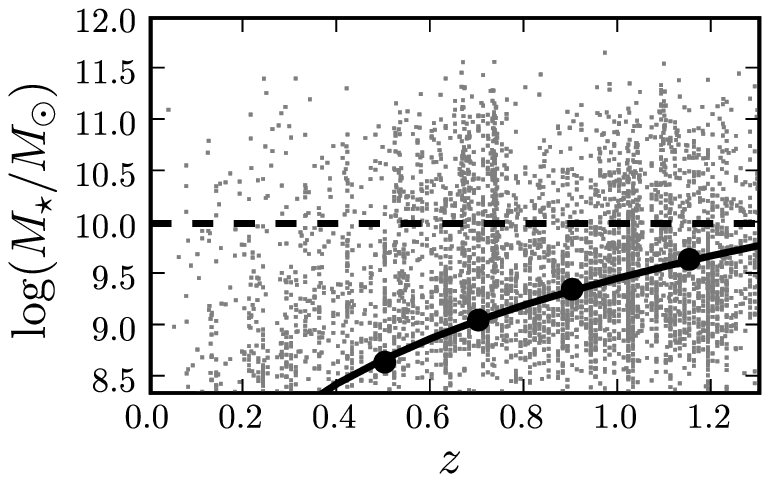}}
\caption{{\it Top}: distribution of $M_B$  vs redshift for IRAC catalogue sources. 
The black dots are the limiting magnitude of the survey at each redshift, defined 
as the third quartile in magnitude distributions. The solid black curve is the best 
fit of the limiting magnitude points by a third-degree polynomial. The black dashed 
line shows the $M_B = -19.5$ limit of our study. {\it Bottom}: distribution of 
$\log (M_{\star}/M_{\odot})$ vs redshift for IRAC catalogue sources. The black 
solid curve shows the stellar mass above which the sample is 75\% complete for 
passively evolving galaxies \citep{pgon08}. The black dashed line shows the 
$\log (M_{\star}/M_{\odot}) = 10$ limit of our study.}
\label{M_vz_z}
\end{figure}

\subsection{Luminosity- and mass-selected samples}\label{bmsample}
The aim of this study is to determine the galaxy merger fraction in $B$-band 
luminosity- and stellar mass-selected samples. The $B$-band study is motivated 
by previous studies, which usually selected their samples in that band. This 
permits us to compare our results with other authors (Sect.~\ref{comparb}). 
Moreover, the stellar mass is a fundamental galaxy property that correlates 
with colour \citep{baldry04} and morphology \citep{conselice06me}.

To determine the luminosity limit in the $B$-band we calculated the third 
quartile of the $M_B$ source distribution at different redshifts, taking 
this as a limiting magnitude \citep[e.g.,][]{pgon08}. In the upper panel of 
Fig.~\ref{M_vz_z} we show $M_B$ vs redshift up to $z_{\rm max} = 1.3$ (grey dots) 
and the limiting magnitude at different redshifts (black bullets). The upper redshift 
limit in our study, $z_{\rm max} = 1.3$, is fixed by the reliability of the asymmetry  
index as a morphological indicator without performing morphological K-corrections 
(see Sect.~\ref{asyrange}, for details). The black solid curve is the least-squares 
fit of the limiting magnitudes by a third-degree polynomial. At redshift $z_{\rm max} 
= 1.3$,  $M_{B,{\rm lim}} \sim  -19.5$, so we selected for our study sources with $M_B \leq -19.5$.

We took as limiting mass at each redshift the stellar mass for which the IRAC 
catalogue is 75\% complete for passively evolving galaxies \citep[see][]{pgon08}. 
In the lower panel of Fig.~\ref{M_vz_z} we show $\log (M_{\star}/M_{\odot})$ vs redshift up to $z_{\rm max} = 1.3$ (grey dots) and the 75\% of completeness at 
different redshifts (black bullets). The black solid curve is the least-squares 
fit of the completeness points by a power-law function. At redshift $z_{\rm max} = 
1.3$, $\log (M_{\star,{\rm lim}}/M_{\odot}) \sim 9.8$, so we selected sources with 
$M_{\star} \geq 10^{10}\ M_{\odot}$ for our study.

\section{ASYMMETRY INDEX}\label{asy}
The automatic asymmetry index ($A$) is one of the CAS morphological indices 
\citep{conselice03}. This index is defined as
\begin{equation}
A = \frac{\sum |I_0 - I_{180}|}{\sum |I_0|} - \frac{\sum |B_0 - B_{180}|}{\sum |I_0|},\label{A}
\end{equation}
where $I_0$ and $B_0$ are the original galaxy and background images, $I_{180}$ and 
$B_{180}$ are the original galaxy and background images rotated 180 degrees, and the 
summation spans all the pixels of the images. The background image is defined in 
detail in the next section. For further details on the asymmetry calculation see 
\citet[][]{conselice00}. This index gives us information over the source distortions 
and we can use it to identify recent merger systems that are highly distorted. In 
previous studies a galaxy was taken to be a recent merger if its asymmetry index 
is $A > A_{\rm m}$, with $A_{\rm m} = 0.35$ \citep[e.g.,][]{conselice03, 
depropris07, bridge07}. This methodology presents several systematic 
effects, such as signal-to-noise dependence \citep{conselice03,conselice05}, 
contamination by non-interacting galaxies with high asymmetry values 
\citep{jogee09,miller08}, contamination by nearby bright sources \citep{depropris07}, 
or the pass-band in which we measure the asymmetry \citep{cassata05,taylor07,conselice08}, 
which must be carefully treated to avoid biased merger fractions. In the following sections we detail how we determined the asymmetry index and its dependence on several factors, 
such as the background image $B_0$ that we use (Sect.~\ref{asyback}), the pass-band in which 
we calculate it (Sects.~\ref{asyrange}, \ref{asyphot}) and the signal-to-noise of the source 
(Sects.~\ref{asyborder}, \ref{asyhomo}, \ref{asyrate}).

\subsection{Asymmetry calculation}\label{asycal}
\subsubsection{Background dependence}\label{asyback}
In Eq.~(\ref{A}) we have a dependence on the background image $B_0$; that is, 
different background images yield different asymmetries for the same source 
\citep{conselice03ff}. To minimize this effect we determined the asymmetry of 
each source with five different background images. These background images are 
sky source-free sections of 50$\times$50 pixels located in the same position in 
the four HST/ACS filter images, and were chosen to span all the GOODS-S area. The 
asymmetry of one source was the median of those five background-dependent asymmetries.

\subsubsection{Pass-bands and redshift range}\label{asyrange}
Galaxy morphology depends on the band of observation \citep[e.g.][]{kuchinski00, 
lauger05, taylor07}. In particular, when galaxies contain both old and young 
populations, morphologies may change very significantly on both sides of the Balmer/4000\AA\ 
break. The asymmetry index limit $A_{\rm m} = 0.35$ was established in the rest-frame 
$B$-band \citep{conselice03}. When dealing with galaxies over a range of redshifts, in 
order to avoid systematic passband biases with redshift, one needs to apply a so-called 
morphological K-correction by performing the asymmetry measurements in a band as close 
as possible to rest-frame $B$ \citep[e.g.,][]{cassata05}, or apply statistical corrections 
for obtaining asymmetries in rest-frame $B$ from asymmetry measurements in rest-frame $U$ 
\citep{conselice08}. Taking advantage of the homogeneous multiband imaging provided by 
the GOODS survey, we entirely avoid morphological K-correction problems in the present
 study by performing asymmetry measurements on all GOODS-S $B_{435}$, $V_{606}$, $i_{775}$, 
and $z_{850}$ images, and using for each source the filter that most closely samples rest-frame $B$.

\begin{figure}[t]
\resizebox{\hsize}{!}{\includegraphics{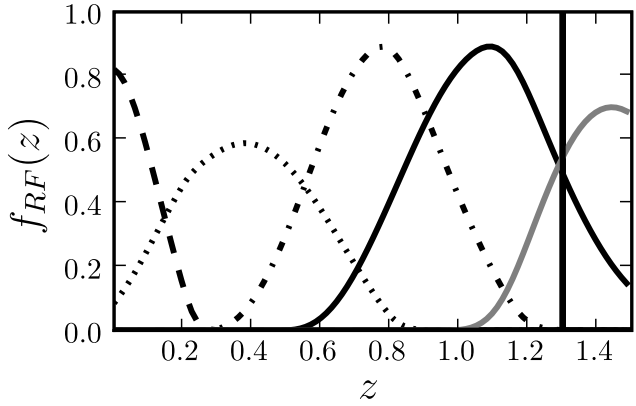}}
\caption{Function $f_{B}(z)$ for the four ACS filters: $B_{435}$ (black dashed curve), 
$V_{606}$ (black dotted curve), $i_{775}$ (black dot-dashed curve), and $z_{850}$ 
(black solid curve). The grey solid curve is the function $f_{U}(z)$ for the $z_{850}$ 
filter. The vertical black solid line is the maximum redshift, $z_{\rm max} = 1.3$, in our study.}
\label{fz_ACS}
\end{figure}

To determine the redshift ranges over which rest $B$-band or $U$-band dominates the 
flux in the four observational HST/ACS filters, $B_{435}$, $V_{606}$, $i_{775}$, and $z_{850}$, we defined the function
\begin{equation}
f_{RF}(z) = 
\frac{\int_0^{\infty}P_{ACS}(\lambda/(1+z)) P_{RF}(\lambda){\rm d}\lambda}{\int_0^{\infty}P_{RF}(\lambda){\rm d}\lambda}, 
\end{equation}
where $P_{RF}$ and $P_{ACS}$ are the transmission curves of the rest-frame reference 
filter and one HST/ACS filter, respectively. In Fig.~\ref{fz_ACS} we show the 
function $f_{B}(z)$ for the four ACS filters (black curves), and $f_{U}(z)$ for
 $z_{850}$ (grey curve). On the basis of this figure, $B_{435}$ asymmetries were
 used for $0 < z \leq 0.15$ sources; $V_{606}$ asymmetries for $0.15 < z \leq 0.55$; 
$i_{775}$ for $0.55 < z \leq 0.9$; and $z_{850}$ for $0.9 < z \leq 1.3$. Staying
 within rest-frame $B$ imposed a maximum redshift of $z_{\rm max} = 1.3$.

Note that, because the ML method used in the merger fraction determination 
(Sect.~\ref{metodo}) takes into account the experimental errors, we had to include 
in the samples not only the sources with $z_i < z_{\rm up}$, where $z_{\rm up}$ is 
the upper redshift in our study, but also sources with $z_i-2\sigma_i < z_{\rm up}$ 
in order to ensure completeness. Because of this, $z_{\rm up}$ must fulfil the 
condition $z_{\rm max} - 2\sigma_{\delta_z}(1 + z_{\rm max}) = z_{\rm up}$, 
which yields $z_{\rm up} \sim 1.1$. We took as minimum redshift in our study 
$z_{\rm min} = 0.1$ because of the lack of sources at lower redshifts. This yields $z_{\rm down} = 
z_{\rm min}+2\sigma_{\delta_z}(1 + z_{\rm min}) \sim 0.2$, which ensures completeness and 
good statistics. Applying these redshift limits we finally have 1740 galaxies with 
$M_B \leq -19.5$ and 982 with $M_{\star} \geq 10^{10}\ M_{\odot}$. The 
number of galaxies quoted here was obtained after removing problematic 
border sources (Sect.~\ref{asyborder}).

\subsubsection{Determining the asymmetry of sources with photometric redshifts}\label{asyphot}
Roughly $\sim 40$\% of the sources in our samples do not have spectroscopic 
redshifts and we rely on photometric redshift determinations. In these cases, our 
source could have its rest-frame $B$-band flux in two observational ACS filters, within 
1$\sigma$. To take this into account we assumed three different redshifts for each 
photometric source: $z_{\rm phot}^{-} = z_{\rm phot} - \sigma_{z_{\rm phot}}$, 
$z_{\rm phot}$, and $z_{\rm phot}^{+} = z_{\rm phot} + \sigma_{z_{\rm phot}}$. 
We determined the asymmetry in these three redshifts. We then performed a weighted 
average of the three asymmetry values such that:
\begin{equation}
A_0 = 0.16A(z_{\rm phot}^{-}) + 0.16A(z_{\rm phot}^{+}) + 0.68A(z_{\rm phot}),\label{proma0}
\end{equation}
where $A(z)$ is the asymmetry of the source at redshift $z$. We used the 
same average procedure with the uncertainties of the three asymmetries and 
added the result in quadrature to the rms of the three asymmetry values to 
obtain $\sigma_{A_0}$. In sources with $z_{\rm spec}$ we only determined the 
asymmetry at the source redshift. L09 show that the two different asymmetry 
determinations do not introduce systematic differences between sources with 
and without spectroscopic information.

\begin{figure}[t]
\resizebox{\hsize}{!}{\includegraphics{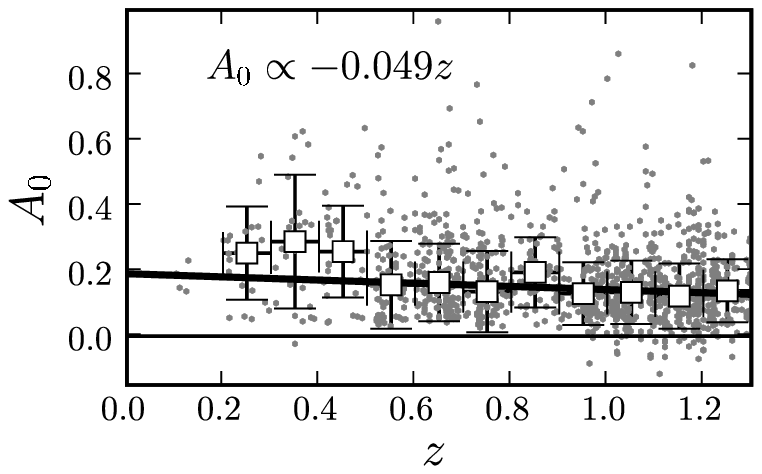}}
\resizebox{\hsize}{!}{\includegraphics{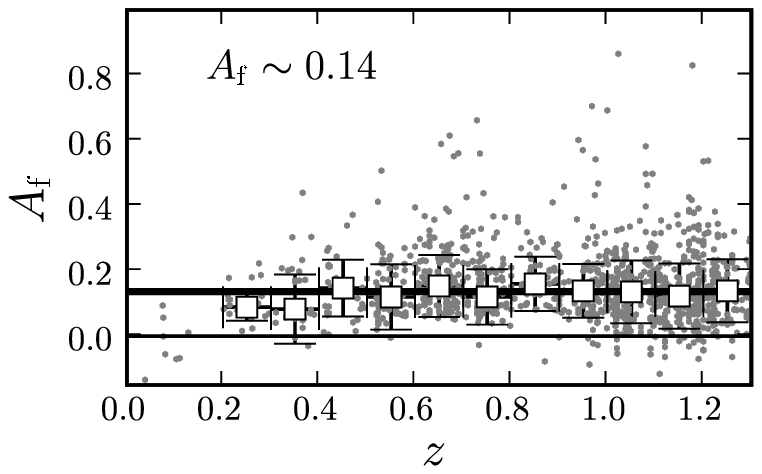}}
\caption{Asymmetry vs redshift in the $M_B \leq -20$ sample (grey dots in both 
panels). {\it Top}: asymmetries of the sources measured on the original images. 
{\it Bottom}: asymmetries of the sources measured on images artificially 
redshifted to $z_{\rm d} = 0.1$. Open squares in both panels are the mean 
asymmetries in 0.1 redshift bins. The black solid line is the least-squares 
linear fit to the mean asymmetries in the [0.5,1.3) redshift interval.}
\label{asyevol}
\end{figure}

\subsubsection{Boundary effects and bright source contamination}\label{asyborder}
The signal-to-noise in HST/ACS decreases near the boundaries of the images, where
 the exposure time is lower. This affects our asymmetry values in two ways: the 
SExtractor segmentation maps that we use to calculate the asymmetry have many 
spurious detections, and any of the five backgrounds defined in Sect.~\ref{asyback} 
is representative of the noisier source background. The problem with segmentation 
maps was noticed previously by \citet{depropris07}, where the segmentation maps for 
50\% of their initial 129 galaxies with $A > 0.35$ are incorrect, or are contaminated 
by bright nearby sources. With this in mind, we visually inspected all the sources 
looking for boundary or contaminated sources. We found that boundary sources had 
systematically high asymmetry values, and had segmentation maps contaminated by 
spurious detections. To avoid biased merger fraction values we excluded all border 
sources (high and low asymmetric) from the samples. We found only two sources 
contaminated by bright nearby sources. For these we redefined the SExtractor 
parameters to construct correct segmentation maps and redetermined the asymmetry.

\subsection{Asymmetries at a Reference Redshift}\label{asyhomo}
The asymmetry index measured on survey images systematically varies with the source 
redshift due, first,  to the $(1+z)^{4}$ cosmological surface brightness dimming, which 
can modify the galaxy area over which asymmetry is measured, and, second, to the loss of 
spatial resolution with $z$. Several papers have attempted to quantify these 
effects by degrading the image spatial resolution and flux to simulate the 
appearance that a given galaxy would have at different redshifts in a given survey. 
\citet{conselice03ff, conselice08}; and \citet{cassata05} degraded a few local 
galaxies to higher redshifts and found that asymmetries decrease with $z$.
 \citet{conselice03ff} also noted that this decrease depends on image depth,
and that luminous galaxies are less affected. In addition, \citet{conselice05} 
show that irregular (high asymmetry) galaxies are more affected than ellipticals 
(low asymmetry). 
A zeroth-order correction for such biases was implemented by 
\citet{conselice03ff,conselice08,conselice09cos} who applied a $\Delta A_z$ term,
 defined as the difference between the asymmetry of local galaxies measured in the 
original images and the asymmetry of the same galaxies in the images degraded to
 redshift $z$. Their final, corrected asymmetries are $A_{\rm f} = A_0 + \Delta A_z$, 
where $A_0$ is the asymmetry measured in the original images. With these corrections, 
all the galaxies have their asymmetry referred to $z = 0$, and the local merger 
criterion $A > A_{\rm m} = 0.35$ is then used.

In their study, L09 improve on the above procedure, and we apply their methodology
 to our data set. We compute a correction term individually for each source in the 
catalogue, but rather than attempting to recover $z=0$ values for $A$ we degrade 
each of the galaxy images to redshift $z_{\rm d} = 1$; we then obtain our final 
asymmetry values $A_{\rm f}$ directly from the degraded images.  With this procedure, 
we take into account that each galaxy is affected differently by the degradation; 
e.g.\ the asymmetry of a low luminosity irregular galaxy dramatically decreases with 
redshift, while a luminous elliptical is slightly affected. We choose $z_{\rm d} = 1$ 
as our reference redshift because a source at this (photometric) redshift
 has $z_{\rm d} + \sigma_{z_{\rm d}} \sim z_{\rm up} = 1.1$; that is, the probability 
that our galaxy belongs to the range of interest is $\sim 85$\%. 
Because we work with asymmetries reduced to $z_{\rm d} = 1$, the asymmetry criterion 
for mergers, $A_{\rm m}$, needs to be reduced to $z = 1$. We discuss this in Sect.~\ref{asyrate}.

We have already mentioned that $\sim$60\% of the sources in the samples have 
spectroscopic redshifts, hence redshift information coming from photometric 
redshifts for the remaining $\sim40$\% of the sources has large uncertainties.
As in the $A_0$ calculation process (Sect.~\ref{asyphot}, Eq.~[\ref{proma0}]), 
to take into account the redshift uncertainty when deriving the asymmetries at 
$z_{\rm d}~=~1$ we started from three different initial redshifts for each 
source, $z_{\rm phot}^{-} = z_{\rm phot} - \sigma_{z_{\rm phot}}$, $z_{\rm phot}$, 
and $z_{\rm phot}^{+} = z_{\rm phot} + \sigma_{z_{\rm phot}}$, and degraded the 
image from these three redshifts to $z_{\rm d} = 1$. We then performed a weighted 
average of the three asymmetry values such that 
\begin{equation}
A_{\rm f} = 0.16A_{1}(z_{\rm phot}^{-}) + 0.16A_{1}(z_{\rm phot}^{+}) + 0.68A_{1}(z_{\rm phot}),\label{promaf}
\end{equation}
where $A_{1}(z)$ denotes the asymmetry measured in the image degraded from 
$z$ to $z_{\rm d} = 1$. When a spectroscopic redshift was available, the 
final asymmetry was simply $A_{\rm f} = A_{1}(z_{\rm spec})$. We did not 
apply any degradation to sources with $z > 1$; that is, we assumed that $A_{1}(z > 1) = A_0$.

To obtain the error of the asymmetry, denoted by $\sigma_{A_{\rm f}}$, 
for sources with photometric redshifts, we averaged the uncertainties of 
the three asymmetries following Eq.~(\ref{promaf}) and added the result in
 quadrature to the rms of the three asymmetry values. The first term 
accounts for the signal-to-noise error in the asymmetry value, while the
 second term is only important when differences between the three asymmetry 
values cannot be explained by the signal-to-noise first term. In sources
 with spectroscopic redshifts we took as $\sigma_{A_{\rm f}}$ the uncertainty
 of the asymmetry $A_{1}(z_{\rm spec})$.

The degradation of the images was performed with \textsc{cosmoshift} 
\citep{balcells03}, which performs repixelation, psf change and flux decrease 
over the sky-subtracted source image. The last \textsc{cosmoshift} step is the 
addition of a random Poisson sky noise to the degraded source image to mimic the 
noise level of the data. As a result of this last step, two \textsc{cosmoshift} 
degradations of the same source yield different asymmetry values. We took the 
asymmetry of each degraded source, $A_{1}(z)$, to be the median of asymmetry 
measurements on five independent degradations of the original source image 
from $z$ to $z_{\rm d} = 1$. With all the aforementioned steps, each $A_1(z)$ 
determination involved 25 asymmetry calculations, while the uncertainty in 
$A_{1}(z)$ was the median of the five individual asymmetry errors. 

The asymmetries $A_{\rm f}$ referred to $z_{\rm d}=1$ provide a homogeneous 
asymmetry set that permits consistent morphological studies in the GOODS-S 
field (L\'opez-Sanjuan et al. 2009b, in preparation). 

\subsection{Asymmetry Trends with Redshift}\label{asyrate}
For a sample of galaxies over a range of redshifts, the statistical change 
with $z$ of the measured asymmetries $A_0$ is the combined effect of loss of
 information (as shown in the previous section) and changes in the galaxy population. 
In contrast, the redshift evolution of $A_{\rm f}$ reflects changes in the galaxy
 population alone, given that the morphological information in the images used to 
determine  $A_{\rm f}$ is homogeneous for the sample. As already discussed in L09 
for the Groth field, we show here that the $z$ trends of $A_0$ and $A_{\rm f}$ 
are quite different.  

In the top panel of Fig.~\ref{asyevol} we show the variation of $A_0$ with redshift 
in a $M_B \leq -20$ selected sample, while in the bottom panel we see the variation 
of $A_{\rm f}$ for the same sample. In both panels, open squares are the median 
asymmetries in $\Delta(z) = 0.1$ redshift bins, and the black solid line is the best 
linear least-squares fit to the $0.5 \leq z < 1.3$ points. $A_0$ is seen to decrease 
with redshift, $A_0 = 0.19 - 0.049z$, while the $A_{\rm f}$ distribution is flat, 
$A_{\rm f} \sim 0.14$. For $A_0$, the negative slope reflects the fact that the 
loss of information with redshift (negative effect on $A$) dominates over genuine 
population variations (a positive effect because galaxies at higher redshift are 
more asymmetric; e.g.\ \citealt{cassata05,conselice05}).In $A_{\rm f}$ the 
information level does not vary with the redshift of the source, so we only see 
population effects. In this case the slope is null, but this is a field-to-field 
effect: L09, with the same methodology and sample selection, obtain $A_f \propto 0.05z$. 
This indicates that we cannot extrapolate results from one field to another, and that 
individual studies of systematics are needed. We take as degradation rate ($\delta_A$) 
the difference between both slopes and assume that the merger condition $A_{\rm m}$ 
varies with redshift as $A_{\rm m}(z) = A_{\rm m}(0) - \delta_A z = 0.35 - \delta_A z$.

\begin{figure}[t]
\resizebox{\hsize}{!}{\includegraphics{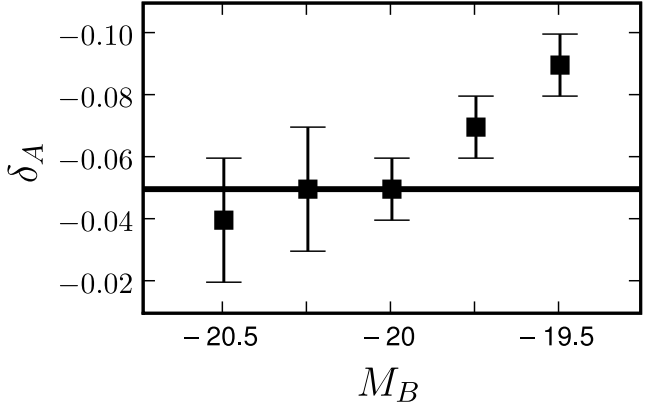}}
\caption{Degradation rate $\delta_A$ vs the selection magnitude $M_B$ of the 
sample (black squares). The black solid line marks $\delta_A = -0.05$, the 
estimated degradation rate for $M_B \leq -20$ galaxies.}
\label{asyrfig}
\end{figure}

\begin{table}
\caption{Degradation rate for different luminosity samples}
\label{asyrtab}
\begin{center}
\begin{tabular}{lcccc}
\hline\hline
Sample selection & $n_{\rm tot}$ & $\delta_A$ & $\overline{A_0}$\\
 & (1) & (2) & (3)\\
\hline
$M_B \leq -19.5$  & 1740 & -0.09 $\pm$ 0.01 &  0.156 \\
$M_B \leq -19.75$ & 1402 & -0.07 $\pm$ 0.01 &  0.160 \\
$M_B \leq -20$    & 1122 & -0.05 $\pm$ 0.01 &  0.163 \\
$M_B \leq -20.25$ & 869  & -0.05 $\pm$ 0.02 &  0.165 \\
$M_B \leq -20.5$  & 648  & -0.04 $\pm$ 0.02 &  0.161 \\
\hline
\end{tabular}
\end{center}
\begin{footnotetext}
TNOTES. Col.~(1) Number of sources in the sample with $0.1~\leq~z~<~1.3$. 
Col.~(2) Degradation rate of the asymmetry, $\Delta A = \delta_A\Delta z$. 
Col.~(3) Median asymmetry of sources with $z < 1$.
\end{footnotetext}
\end{table}

Is the degradation rate the same for all luminosity selections? 
We expect less asymmetry variation with redshift in bright samples,
because they are less affected by cosmological dimming \citep{conselice03}.
 We repeated the previous analysis with different $M_B$ selection cuts, 
from $M_B \leq -20.5$ to $M_B \leq -19.5$ (the latter is the limiting 
magnitude in our study, Sect.~\ref{bmsample}). We summarize the results in 
Table~\ref{asyrtab} and Fig.~\ref{asyrfig}: asymmetry is more affected by
 redshift changes in less luminous samples, as expected. Interestingly, the
 degradation rate is roughly constant up to $M_B = -20$, $\delta_A 
\sim -0.05$ (black solid line in Fig.~\ref{asyrfig}), but then becomes 
more pronounced by a factor of 2, $\delta_A \sim -0.09$, in only 0.5 
magnitudes. One could argue that the sharp increase of $\delta_A$ for 
samples including $M_B > -20$ sources arises because such sources have 
higher initial asymmetry $A_0$: a faint irregular galaxy is more affected 
by loss of information than a bright elliptical. However, we see in the 
last column of Table~\ref{asyrtab} that the mean asymmetry of sources with 
$z < 1.0$ is similar in all samples, $\overline{A_0} \sim 0.16$. Hence, 
the degradation rate increases because faint sources have lower signal-to-noise
 than luminous ones. Because of this, we decided to restrict our study to the 
1122 sources with $M_B \leq -20$ to ensure that degradation affects all the galaxies 
in our sample in the same way, making the merger condition $A_{\rm m}(1) = 0.35 - 
\delta_A = 0.30$ representative.
 
\begin{figure}[t]
\resizebox{\hsize}{!}{\includegraphics{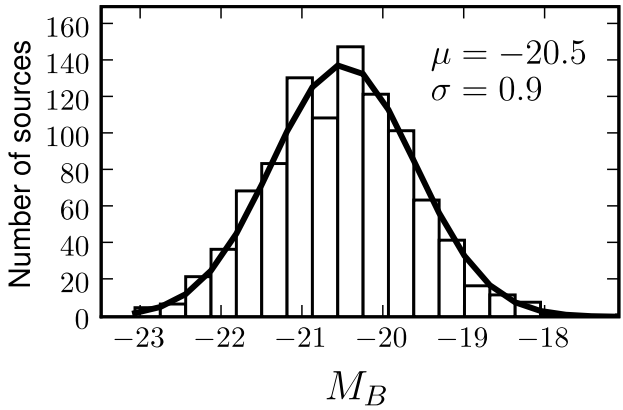}}
\caption{$M_B$ distribution of $M_{\star} \geq 10^{10}\ M_{\odot}$
 galaxies. The black solid line is a Gaussian with $\mu = -20.5$ and $\sigma = 0.9$.}
\label{magmass}
\end{figure}
 
How important is this luminosity dependence for the mass-selected sample? 
The $M_B$ distribution of $M_{\star} \geq 10^{10}\ M_{\odot}$ galaxies is
 well described by a Gaussian with $\mu = -20.5$ and $\sigma = 0.9$,
 Fig.~\ref{magmass}. We found that 70\% of the galaxies have $M_B \leq -20$, and 
that the degradation rate for the whole sample is $\delta_A = -0.05$. This tells 
us that the faint sources in this sample do not significantly affect the 
degradation rate, making the $M_B \leq -20$ merger condition representative 
also for the mass-selected sample. In conclusion, we used $A_{\rm m}(1) = 0.30$ for both samples.

\section{MERGER FRACTION DETERMINATION}\label{metodo}
Following \citet[][]{conselice06ff}, the merger fraction by morphological criteria is
\begin{equation}\label{fmg}
f^{\rm mph} = \frac{\kappa \cdot n_{\rm m}}{n_{\rm tot} + (\kappa - 1) n_{\rm m}},
\end{equation}
where $n_{\rm m}$ is the number of the distorted sources with $A > A_{\rm m}$, 
and $n_{\rm tot}$ is the total number of sources in the sample. If $\kappa \geq 
2$ we obtain the galaxy merger fraction, $f_{\rm gm}^{\rm mph}$, the fraction of
 galaxies undergoing mergers, and $\kappa$ represents the average number of galaxies 
that merged to produce one distorted remnant. If $\kappa = 1$ we obtain the merger
 fraction, $f_{\rm m}^{\rm mph}$: the number of merger events in the sample. We
 use $\kappa = 1$ throughout this paper.

The steps we followed to obtain the merger fraction are described in detail in 
LGB08. In this section we provide a short summary. If we define a 
two-dimensional histogram in the redshift--asymmetry space and normalize this 
histogram to unity, we obtain a two-dimensional probability distribution defined
 by the probability of having one source in bin $[z_k, z_{k+1}) \cap [A_l, A_{l+1})$,
 namely $p_{kl}$, where the index $k$ spans the redshift bins of size $\Delta z$, and
 the index $l$ spans the asymmetry bins of size $\Delta A$. We consider only two 
asymmetry bins split at $A_{\rm m}$, such that the probabilities $p_{k1}$ 
describe highly distorted galaxies (i.e.\ merger systems), while the probabilities
 $p_{k0}$ describe normal galaxies. With those definitions, the morphologically based
 merger fraction in the redshift interval $[z_k, z_{k+1})$ becomes
\begin{equation}\label{ffpkl}
f_{{\rm m},k}^{\rm mph} = \frac{p_{k1}}{p_{k0}+p_{k1}}.
\end{equation}
In LGB08 they describe a maximum likelihood (ML) method that yields the most 
probable values of $p_{kl}$ taking into account not only the $z$ and $A$ values, 
but also their experimental errors. The method is based on the minimization of the 
joint likelihood function, which in our case is
\begin{eqnarray}
L(z_{i},A_{i}|p^{\prime}_{kl},\sigma_{z_{i}},\sigma_{A_{i}})\nonumber
\end{eqnarray}
\begin{equation}
= \sum_i \biggr[ \ln \bigg\{ \sum_k\sum_l \frac{{\rm e}^{p'_{kl}}}{4}{\rm ERF}(z,i,k){\rm ERF}(A,i,l) 
\bigg\}\biggr]\label{MLfunc},
\end{equation}
where
\begin{equation}
{\rm ERF}(\eta,i,k) \equiv {\rm erf}\bigg(\frac{\eta_{i} - 
\eta_{k+1}}{\sqrt{2} \sigma_{\eta_{i}}}\bigg) - {\rm erf}\bigg(\frac{\eta_{i} -
 \eta_{k}}{\sqrt{2} \sigma_{\eta_{i}}}\bigg).\label{ERF}
\end{equation}
In the above equations, ${\rm erf}(x)$ is the error function; $z_i$ and $A_i$ are 
the redshift and asymmetry values of source $i$, respectively; $\sigma_{z_i}$ and 
$\sigma_{A_i}$ are the observational errors in redshift and asymmetry of source $i$,
 respectively; and the new variables $p'_{kl} \equiv \ln (p_{kl})$ are chosen to avoid
 negative probabilities. Equation~(\ref{MLfunc}) was obtained by assuming that the real 
distribution of galaxies in the redshift--asymmetry space is described by a two-dimensional 
distribution $p_{kl} \equiv \exp(p'_{kl})$, and that the experimental errors are Gaussian. 
Note that changing variables to $p'_{kl} = \ln(p_{kl})$, Eq.~(\ref{ffpkl}) becomes
\begin{equation}\label{ffpklp}
f_{{\rm m},k}^{\rm mph} = \frac{{\rm e}^{p'_{k1}}}{{\rm e}^{p'_{k0}}+{\rm e}^{p'_{k1}}}.
\end{equation}
LGB08 show, using synthetic catalogues, that the experimental errors tend to smooth an
 initial two-dimensional distribution described by $p_{kl}$, due to spill-over of 
sources to neighbouring bins.  This leads to a $\sim10-30$\% overestimate of the galaxy 
merger fraction in typical observational cases. L09 and \citet{lotz08ff} find similar
 trends in their study of the morphological merger fraction in the Groth Strip. LGB08 
additionally show that, thanks to the use of the ML method, they can accurately recover
the initial two-dimensional distribution: the fractional difference between 
the input and ML method merger fractions 
is a tiny $\sim 1$\% even when the experimental errors are similar to the bin size. 
That is, the ML results are not biased by the spill-over of sources to neighbouring bins.

We obtained the morphological merger fraction by applying 
Eq.~(\ref{ffpklp}) using the probabilities $p'_{kl}$ recovered by the ML 
method. In addition, the ML method provides an estimate of the 68\% confidence 
intervals of the probabilities $p'_{kl}$, which we use to obtain the $f_{{\rm m},k}^{\rm mph}$ 
68\% confidence interval, denoted $[\sigma^{-}_{f_{{\rm m},k}^{\rm mph}}, 
\sigma^{+}_{f_{{\rm m},k}^{\rm mph}}]$. This interval is asymmetric because 
$f_{{\rm m},k}^{\rm mph}$ is described by a log--normal distribution due to the 
calculation process (see LGB08 for details). Note that, in LGB08, $\kappa = 2$ is 
used in Eq.~(\ref{fmg}), but the method is valid for any $\kappa$ value.

We also determined the morphological merger fraction by classical counts, 
$f_{\rm m, class}^{\rm mph} = n_{\rm m}^{\rm class} / n_{\rm tot}^{\rm class}$, 
where $n_{\rm m}^{\rm class}$ is the number of galaxies in a given bin with 
$A_{\rm f} > A_{\rm}$, and $n_{\rm tot}^{\rm class}$ is the total number of 
sources in the same bin. We obtained the $f_{\rm m, class}^{\rm mph}$ uncertainties 
assuming Poissonian errors in the variables.

Finally, and following L09, sect.~4.1, we performed simulations with synthetic 
catalogues to determine the optimum binning in redshift for which the ML method 
results are reliable. The simulations were made in the same way as in L09, so here 
we only report the results of the study: we can define up to three redshift bins, 
namely $z_1$ = $[0.2,0.6)$, $z_2 = [0.6,0.85)$, and $z_3 = [0.85,1.1)$. The first 
bin is wider than the other two, 0.4 vs 0.25, because of the lower number of sources 
in the first interval. In the next section we study the merger fraction evolution 
with redshift with these three bins (\S~\ref{ffevolz}). We will also provide statistics 
for the $z_0$ = $[0.2,1.1)$ bin in order to compare the ML and classical merger 
fraction determinations.
 
\begin{table*}
\caption{Sample characteristics in the $0.2 \leq z < 1.1$ range}
\label{sampletab}
\begin{center}
\begin{tabular}{lccccccccc}
\hline\hline
Sample selection & $n_{\rm tot}^{\rm class}$ & $n^{\rm class}_{\rm m}$ & $f_{\rm m,class}^{\rm mph}$ & 
$n_{\rm tot}^{\rm ML}$ & $n^{\rm ML}_{\rm m}$ & $f_{\rm m,ML}^{\rm mph}$\\
 & (1) & (2) & (3) & (4) & (5) & (6)\\
\hline
$M_B \leq -20$                       & 793 & 61 & $0.077 \pm 0.010$ & 881.9 & 30.7 & $0.035^{+0.010}_{-0.008}$\\
$M_{\star} \geq  10^{10}\ M_{\odot}$ & 759 & 38 & $0.050 \pm 0.008$ & 819.3 & 20.2 & $0.025^{+0.008}_{-0.006}$\\
\hline
\end{tabular}
\end{center}
\begin{footnotetext}
TNOTES. Col. (1) Number of galaxies with $0.2 \leq z < 1.1$ by classical counts. Col. (2)
 Number of distorted galaxies with $0.2 \leq z < 1.1$ and $A_{\rm f} > 0.30$ by classical
 counts. Col. (3) Morphological major merger fraction by classical counts. Col. (4) Number 
of galaxies with $0.2 \leq z < 1.1$ by ML method. Col. (5) Number of distorted galaxies 
with $0.2 \leq z < 1.1$ and $A_{\rm f} > 0.30$ by ML method. Col. (6) Morphological major 
merger fraction by ML method.
\end{footnotetext}
\end{table*}

\begin{table*}
\caption{Morphological major merger fractions $f_{\rm m}^{\rm mph}$ in GOODS-S}
\label{fftab}
\begin{center}
\begin{tabular}{lccccc}
\hline\hline
Sample selection & z = 0.4 & z = 0.725 & z = 0.975 & $f_{\rm m}^{\rm mph}(0)\ ^{\mathrm{a}}$ & $m\ ^{\mathrm{a}}$\\
\hline
$M_B \leq -20$  &	$0.023^{+0.022}_{-0.011}$ & $0.031^{+0.016}_{-0.011}$ & 
$0.043^{+0.015}_{-0.011}$ & $0.013 \pm 0.003\ ^{\mathrm{b}}$ & $1.8 \pm 0.5\ ^{\mathrm{b}}$\\
$M_{\star} \geq  10^{10}\ M_{\odot}$ & $0.006^{+0.018}_{-0.005}$ 
& $0.022^{+0.013}_{-0.008}$ & $0.037^{+0.016}_{-0.011}$ & $(1.0 \pm 0.2) \times 10^{-3}$ & $5.4 \pm 0.4$\\
\hline
\end{tabular}
\end{center}
\begin{list}{}{}
\item[$^{\mathrm{a}}$]Best $f_{\rm m}^{\rm mph}(z) = f_{\rm m}^{\rm mph}(0)(1+z)^m$ fit to the data.
\item[$^{\mathrm{b}}$]This fit includes the \cite{depropris07} local value.
\end{list}
\end{table*}

\begin{figure}[t]
\resizebox{\hsize}{!}{\includegraphics{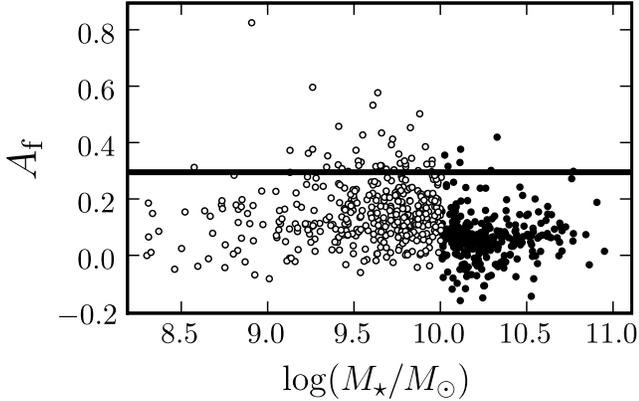}}
\caption{Asymmetry vs $\log (M_{\star}/M_{\odot})$ for {\it light-weight} 
(open circles) and {\it faint} (bullets) samples (see text for details). 
The black solid line shows the merger criterion $A_{\rm m}(1) = 0.30.$}
\label{mbmass}
\end{figure}

\section{RESULTS}\label{results}
We summarize in Table~\ref{sampletab} the main characteristics of
 the two samples under study; i.e.\ the total ($n_{\rm tot}$) and 
distorted ($n_{\rm m}$) number of sources, both for classical counts 
($n^{\rm class}$) and the ML method ($n^{\rm ML}$), and major merger 
fractions. Note that the number of ML method galaxies is not an integer.
 Indeed, the ML method gives us a statistical estimate of the probability 
$p_{kl} = \exp(p'_{kl})$ of finding one source in the redshift bin $k$, and 
in the asymmetry bin $l$, so the estimated number of galaxies in that bin,
 $n_{kl,{\rm ML}} = N_{\rm tot}p_{kl}\Delta z \Delta A$, where $N_{\rm tot}$ 
is the total number of galaxies in the sample, need not be an integer. 
The merger fraction by the ML method is roughly half that in the classical determination 
(0.035 vs 0.077 in the luminosity-selected sample, 0.025 vs 0.050 in the 
mass-selected sample). This highlights the fact that, whenever the spill-over 
effect of large measurement errors is not taken into account, morphological merger 
fractions can be overestimated by a factor of $\sim 2$. We use this result later 
in Sect.~\ref{ffhighz}, and in the next section we use only merger fractions 
obtained by the ML method.

We find that correction of redshift-dependent biases is equally important.  If we use the raw asymmetry values determined on the original images, and apply the local Universe merger selection criterion $A_0 > 0.35$, the resulting merger fractions come up a factor 2 higher than the ones listed in Table~\ref{sampletab}.  Recall that the latter come from $A_{\rm f}$ values homogenised to a common reference $z_{\rm d} = 1$ (Sect.~\ref{asyhomo}).  This emphasises that published merger fractions which do not work with redshift-homogeneous data, may be significantly biased.  Interestingly, an identical comparison to the one just described, applied to Groth strip data, lead L09 to conclude that redshift effects are \textsl{not} important for merger fraction determinations.  The different behaviour of the Groth data from L09 and our GOODS-S data might be due to cosmic variance, or to depth differences between the two data sets.  In general though, artificial redshifting of the galaxies is needed to ensure reliable results. 


Table~\ref{sampletab} shows that the merger fraction from the mass-selected sample is 
lower than that from the luminosity-selected sample. What is the origin of this 
difference? To answer this question, we define two subsamples: the {\it faint} sample 
(galaxies with $M_B > -20$ and $M_{\star} \geq  10^{10}\ M_{\odot}$), and the {\it 
light-weight} sample (sources with $M_{\star} <  10^{10}\ M_{\odot}$ and $M_B \leq -20$). 
The {\it faint} sample comprises 272 sources, while the {\it light-weight} sample
 comprises 408 sources. In Fig.~\ref{mbmass} we show both samples in the mass--asymmetry 
plane: light-weight galaxies have higher asymmetry, $\overline{A_{\rm f}} = 0.14$, while 
faint galaxies are more symmetric, $\overline{A_{\rm f}} = 0.07$. The 
{\it light-weight} sample comprises 43 sources with $A_{\rm f} > 0.30$ (10.5\% of the sample), 
while the {\it faint} sample comprises only seven distorted sources (2.5\% of the sample). 
These numbers suggest, in agreement with L09, that: (i) an important fraction of the 
$B$-band high asymmetric sources are low-mass disc--disc merger systems that, due to 
merger-triggered star-formation, have their $B$-band luminosity boosted by 1.5 magnitudes 
\citep{bekki01}, enough to fulfil our selection cut $M_B \leq -20$; and (ii) the 
faint objects are earlier types dominated by a spheroidal component which, when subject
 to a major merger, does not distort enough to be picked up as merger systems by our 
asymmetry criterion.

\begin{table*}
\caption{Morphological merger fraction in GOODS-S at $0.6 \leq z < 0.85$}
\label{lsstab}
\begin{center}
\begin{tabular}{lcccc}
\hline\hline
Sample selection & $n_{\rm LSS}$ & w/ LSS & w/o LSS & LSS ($z = 0.735$)\\
 & (1) & (2) & (3) & (4)\\
\hline
$M_B \leq -20$  & 72 & $0.031^{+0.016}_{-0.011}$ & $0.026^{+0.020}_{-0.011}$ & $0.044^{+0.032}_{-0.018}$ \\
$M_{\star} \geq  10^{10}\ M_{\odot}$ & 94 & $0.022^{+0.013}_{-0.008}$ & 
$0.018^{+0.016}_{-0.008}$ & $0.032^{+0.023}_{-0.014}$\\
\hline
\end{tabular}
\end{center}
\begin{footnotetext}
TNOTES. Col. (1) Number of galaxies in the Large Scale Structure (LSS). Col. 
(2) Merger fraction in the sample {\it with LSS}. Col. (3) Merger fraction in 
the sample {\it without LSS}. Col. (4) Merger fraction in the LSS.
\end{footnotetext}
\end{table*}

\subsection{Merger fraction evolution}\label{ffevolz}
We summarize in Table~\ref{fftab} the morphological merger fraction at different 
redshifts in GOODS-S. We obtain low merger fractions, always lower than 0.06, 
similar to the L09 results for the Groth field. The merger fraction increases with 
redshift in both the luminosity- and the mass-selected samples, but this growth is 
more prominent in the mass-selected sample. We can parameterize the merger fraction 
evolution as
\begin{equation}\label{fffit}
f_{\rm m}^{\rm mph}(z) = f_{\rm m}^{\rm mph}(0) (1+z)^m
\end{equation}
and fit our 
data. Note that, in the luminosity-selected sample, we also use the $M_B \leq -20$ 
estimation from L09 of the $M_B \lesssim -19$ local merger fraction, drawn from the 
MGC\footnote{www.eso.org/~jliske/mgc} (Millenium Galaxy Catalogue), 
from \citet{depropris07}: $f_{\rm m}^{\rm mph}(0.07) = 0.014^{+0.003}_{-0.003}$.
We summarize the results in Table~\ref{fftab} and Fig.~\ref{fffig}. The merger index
 $m$ is higher ($3\sigma$) in the mass-selected sample (bullets) than in the 
luminosity-selected sample (open triangle for \citealt{depropris07} local value; 
open squares for our data), 5.4 vs 1.8, while the merger fraction in the local universe 
is lower in the mass-selected sample, 0.001 vs 0.013. The fact that the higher $m$, the 
lower $f_{\rm m}^{\rm mph}(0)$, was predicted by semianalytical models \citep{khochfar01}. 
We compare these values with those from previous studies in Sect.~\ref{comparb}.

\subsection{Large Scale Structure effect}\label{lss}
It is well known that the more prominent large scale structure (LSS) in the 
GOODS-S field is located at redshift $z = 0.735$ \citep{ravikumar07}. In order 
to check the effect of this LSS on our derived merger fractions, we recalculated 
them by excluding the sources within 
$\delta v \leq 1500\ {\rm km\ s}^{-1}\ (\delta z \sim 0.01)$ of $z = 0.735$ \citep{rawat08}. 
In Table~\ref{lsstab} we summarize the number of sources in the LSS for each sample
 ($n_{\rm LSS}$), and the previous and recalculated merger fractions, both in the
 field and in the structure. The merger fraction is higher in the LSS than in the
 field. Note that the variation in the field values is well reported by the error bars. 
How does this LSS affect the previously inferred merger evolution? If we again fit the 
data without LSS, we find that $f_{\rm m}^{\rm mph}(0)$ does not change, while the value 
of $m$ decreases only by 0.1 in both the luminosity- and the mass-selected samples, so
our conclusions remain the same. We shall therefore use the fit values in Table~\ref{fftab} 
in the remainder of the paper. We concentrate on the LSS at $z = 0.735$, and ignore other structures in GOODS-S. The next two more important ones are located at $z = 0.66$ 
and $z = 1.1$. The former is an overdensity in redshift space, but not in the sky plane, 
while the latter is a cluster, but comprises an order of magnitude fewer sources than the
 $z = 0.735$ structure (145 vs 12, \citealt{Adami05}).

\section{DISCUSSION}\label{discussion}
\begin{figure*}[!t]
\resizebox{\hsize}{!}{\includegraphics{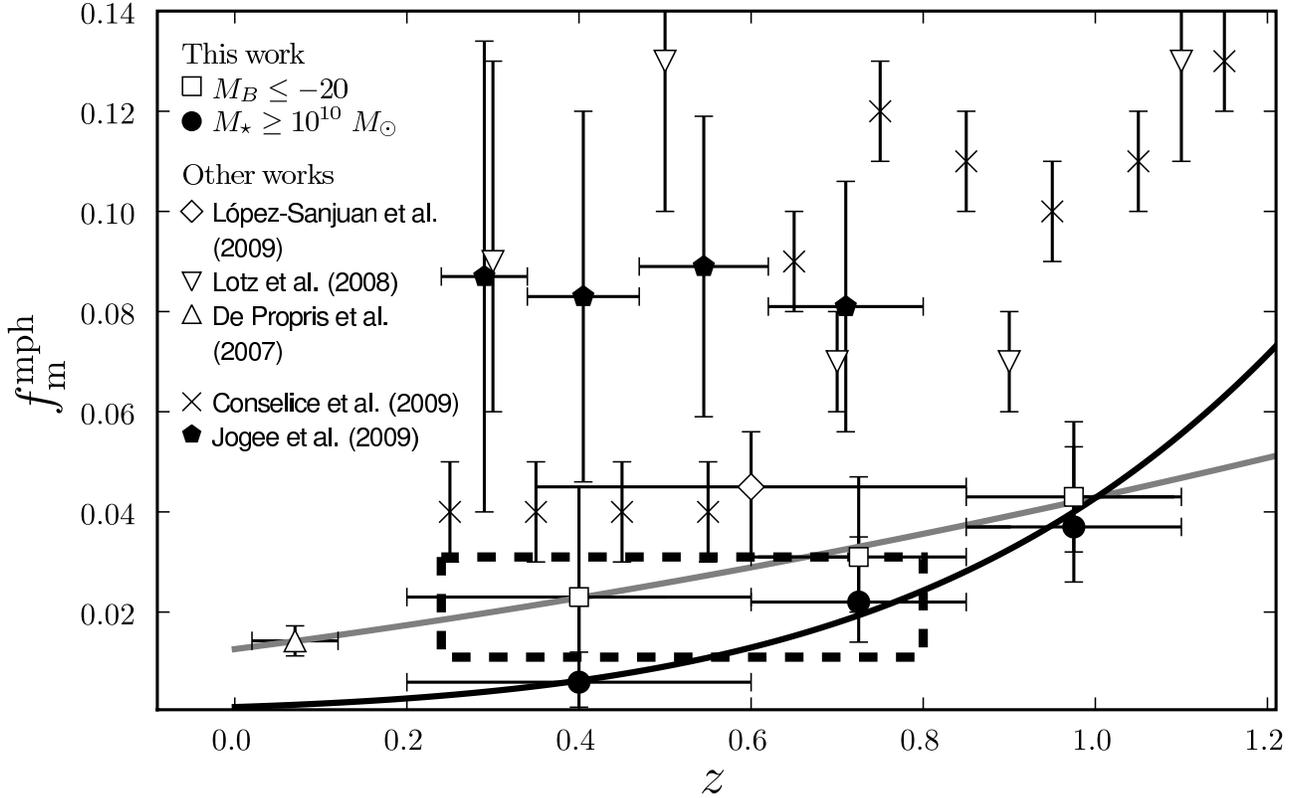}}
\caption{Morphological merger fraction vs redshift for $M_B \leq -20$ (open squares) 
and $M_{\star} \geq 10^{10}\ M_{\odot}$ galaxies (bullets). The error bars do {\it not} include cosmic variance (Sec.~\ref{cosvar}). The grey/black solid lines are the 
least-squares fit of $f_{\rm m}^{\rm mph}(z) = f_{\rm m}^{\rm mph}(0) (1+z)^m$ to 
the data in the luminous/mass case, respectively. The open triangle is the
$M_B \leq -20$ estimate by L09 of the \citet{depropris07} $M_B \lesssim -19$ merger
 fraction, open inverted triangles are from \citet{lotz08ff}, open diamond is for
 $M_B \leq -20$ galaxies in Groth strip from L09, crosses are for 
$M_{\star} \geq 10^{10}\ M_{\odot}$ galaxies from \citet{conselice09cos}, 
and filled pentagons are minor+major mergers for $M_{\star} \geq 5 \times 10^{10}\ M_{\odot}$ 
galaxies from \citet{jogee09}. The dashed lines marks the major merger fraction expected by
 \citet{jogee09}.}
\label{fffig}
\end{figure*}

First we compare our results with merger fraction determinations from other authors. 
In Fig.~\ref{fffig} we show our results (open squares for $M_B\leq -20$ galaxies and
 bullets for $M_{\star} \geq 10^{10}\ M_{\odot}$ galaxies). The other points are those
 from  the literature: the $M_B \leq -20$ estimate by L09 of the \citet{depropris07} 
$M_B \lesssim -19$ merger fraction; the merger fraction for $B$-band luminosity selected 
galaxies in AEGIS\footnote{http://aegis.ucolick.org/} (All-Wavelength Extended Groth Strip
 International Survey) from \citet{lotz08ff}; the results from \citet{conselice09cos} in 
COSMOS\footnote{http://cosmos.astro.caltech.edu/index.html} (Cosmological Evolution Survey) 
and AEGIS for $M_{\star} \geq 10^{10}\ M_{\odot}$ galaxies; and the merger fraction for 
$M_{\star} \geq 5 \times 10^{10}\ M_{\odot}$ galaxies in 
GEMS\footnote{http://www.mpia.de/GEMS/gems.htm} (Galaxy Evolution from 
Morphology and SEDs) from \citet{jogee09}. Note that the mass selection from 
\citet{jogee09} has been adapted to a Salpeter IMF \citep{salpeter55}. All the 
previous merger fractions except those from \citet{jogee09} are from automatic indices for major mergers. The \citet{jogee09} results are by visual morphology and reflect
 major+minor mergers; the dashed rectangle marks their expected major merger fraction. 
For luminosity-selected samples (open symbols) our values are in good agreement with
 \citet{depropris07}, but are lower than those from \citet{lotz08ff}, who apply different 
sample selection and merger criteria from ours and do not correct the effect of observational
 errors, thus making comparison difficult.

In the mass-selected case our results are in good agreement with the expected 
visual major merger fraction from \citet{jogee09} (dashed lines), supporting 
the robustness of our methodology for obtaining major merger fractions statistically. 
Our values are significantly lower that those of \citet{conselice09cos}, especially at $z 
\gtrsim 0.7$, where there is a factor 3 difference. The asymmetry calculation performed
 by \citet{conselice09cos}  does not take into account the spill-over effect of observational errors in their merger fraction 
determination.  We show here that such effects may lead to the higher value obtained by them.  
\citet{conselice09cos} assume two main statistical corrections at $z \gtrsim 0.7$: the information degradation bias ($\Delta A_{z}$, \S~\ref{asyhomo}) and the morphological 
K-correction \citep[$\Delta A_{K}$, see][for details]{conselice08}. The first correction 
is $\Delta A_{z} = 0.5$ and has an associated uncertainty of 
$\sigma_{\Delta A_z} \sim 0.08$ \citep[][Table~1]{conselice03ff}. The morphological
 K-correction depends on redshift; to simplify the argument, we do not consider its 
uncertainty in the following. In addition, each source asymmetry has its own signal-to-noise 
uncertainty, which in our study is $\sim 0.03$ at these redshifts.  
We reproduced the same methodology applied by \citet{conselice09cos} on synthetic catalogs created as in Sect.~\ref{metodo}.  For further details about simulation parameters and assumptions, 
see L09. In the simulations we defined two redshift intervals, namely $z_2 = [0.6,0.85)$ 
and $z_3 = [0.85,1.1)$, taking our results in these redshift intervals as input merger 
fractions, 
$f_{\rm m}^{\rm mph} = 0.022$ in the first interval, and $f_{\rm m}^{\rm mph} = 0.037$ in the second. We then extracted 2000 random sources in the redshift-asymmetry plane, 
applying an asymmetry error to them of $\sigma_{A} = 0.08$, which is representative of 
the asymmetry uncertainties in \citet{conselice09cos}. We assumed $\sigma_{z} = 0$ for simplicity.  Merger fractions were derived from classical histograms as in \citet{conselice09cos}. We 
repeated this process 100 times and averaged the results.  This process yields $f_{\rm m}^{\rm mph} 
\sim 0.11$ in the first interval, and $f_{\rm m}^{\rm mph} \sim 0.13$ in the second, 
which is similar to \citet{conselice09cos} results at these redshifts. In contrast,  
the ML method was able to recover the input merger fractions.  The exercise demonstrates that the observed differences 
betwen the two studies can be naturally explained as a bias introduced in \citet{conselice09cos} by not accounting for spill-over of sources due to observational errors.  
The fact that 
\citet{conselice09cos} study is performed over $\sim 20000$ galaxies, 20 times more sources 
than in our study, cannot correct the errors.  As emphasized by LGB08, experimental systematic errors are not cured by increasing sample size: the ML method is needed.

\subsection{Groth vs GOODS-S merger fractions: cosmic variance effect\label{cosvar}}
L09 report a morphological merger fraction 
\begin{equation}
f_{\rm m, GS}^{\rm mph}(z = 0.6, M_B \leq -20) = 0.045^{+0.014}_{-0.011}
\end{equation}
in the Groth field (open diamond in Fig.~\ref{fffig}). How does this 
value compare with the one obtained in GOODS-S? If we use the same 
selection as in L09, this is, $M_B \leq -20$ galaxies with $0.35 \leq 
z < 0.85$, the major merger fraction in GOODS-S is
\begin{equation}
f_{\rm m, GOODS}^{\rm mph}(z = 0.6, M_B \leq -20) = 0.032^{+0.013}_{-0.009}.
\end{equation}
We can see that both values are consistent within their errors. Because 
both values are determined using the same methodology and sample selection, 
the difference of $\Delta f_{\rm m} = 0.013$ may be explained by cosmic 
variance, denoted by $\sigma_{\rm v}$. Following \citet{somerville04}, we 
infer that the effect of cosmic variance for the typical merger density 
($\sim 10^{-4}\ {\rm Mpc}^{-3}$, Sect.~\ref{mrgoods}) and GOODS-S/Groth 
volume is $\sim 60$\%. That is, we expect $\sigma_{\rm v} \sim 0.027$ in 
Groth and $\sigma_{\rm v} \sim 0.019$ in GOODS-S, so the difference between 
both merger fraction determinations can indeed be explained as cosmic variance 
effect. Averaging both values, the morphological merger fraction at $z = 0.6$ is
\begin{equation}
f_{\rm m}^{\rm mph}(z = 0.6, M_B \leq -20) = 0.038 \pm 0.012,
\end{equation}
where the error is the expected $\sigma_{\rm v} \sim 30$\% due to combining 
two separate fields \citep[see][for details]{somerville04}.

\subsection{Morphological merger fraction evolution in previous studies}\label{comparb}
In Sect.~\ref{ffevolz} we obtained the values of $m$ and 
$f_{\rm m}^{\rm mph}(0)$ that describe the morphological merger fraction 
evolution in GOODS-S. In this section we compare these values with those in the 
literature, where morphological works in $B$-band selected samples are common. 
L09 study the merger fraction for $M_B\leq -20$ galaxies in Groth by asymmetries
 and taking into account the experimental error bias. Combining their results with 
the literature, they obtain $m = 2.9 \pm 0.8$, consistent to within $\sim 1\sigma$ 
with our result. \citet{lotz08ff} study the merger fraction in an
 $M_B\leq -18.83 - 1.3z$ selected sample by $G$ and $M_{20}$ morphological 
indices. Their results alone suggest $m = 0.23 \pm 1.03$, but when combined with 
others in the literature they obtain $m = 2.09 \pm 0.55$. The first case does not 
match the local morphological merger fraction by \citet{depropris07}: with a 
similar luminosity cut, $M_B \lesssim -19$, and taking into account the different
 methodologies (see L09, for details), the merger fractions are very different, 0.006 
\citep{depropris07} vs 0.07 \citep{lotz08ff}. Because of this, the second $m$ value is 
preferred. \citet{kamp07} study the fraction of visually distorted galaxies in 
SDSS\footnote{http://www.sdss.org/} (Sloan Digital Sky Survey, local value) and COSMOS
 ($z \sim 0.7$ value) for $M_B \leq -19.15$ galaxies. They find that $m = 3.8 \pm 1.8$,
 higher than our value, but consistent to within $\sim 1\sigma$. Finally, 
\citet{conselice03ff} study the morphological merger fraction of $M_B \leq -20$ by 
asymmetries. However, due to the small area of their survey, they have high uncertainties 
in the merger fraction at $z \lesssim 1$, so we do not compare our results with theirs. 
In summary, the morphological major merger fraction evolution in $M_B$ samples up to
 $z \sim 1$ is consistent with a $m = 2.2 \pm 0.4$ evolution (weighted average of the
 previous $m$ values), although more studies are needed to understand its dependence 
on different luminosity selections. 

The only previous morphological merger fractions in  $M_{\star} \geq 10^{10}\ M_{\odot}$ 
selected samples are from \citet{conselice03ff,conselice08,conselice09cos}. The small
 areas in the first two studies (HDF\footnote{http://www.stsci.edu/ftp/science/hdf/hdf.html} 
in \citealt{conselice03ff} and UDF\footnote{http://www.stsci.edu/hst/udf} in
 \citealt{conselice08}) make their $z \lesssim 1$ values highly undetermined, 
and we use their $z \gtrsim 1$ values to constrain the merger fraction evolution at 
higher redshifts in Sect.~\ref{ffhighz}. \citet{conselice09cos} find $m = 3.8 \pm 0.2$. 
This value is lower than ours, but it is higher than typical values in $B$-band studies, 
supporting the hypothesis that merger fraction evolution in mass-selected samples is more 
important than in luminosity-selected samples.

Other asymmetry studies have used different selection criteria from ours: 
\citet{cassata05} obtain a merger fraction evolution $m = 2.2 \pm 0.3$ in an $m_{K_{\rm s}}
 < 20$ selected sample, and combining their results with others in the literature. 
\citet{bridge07} perform their asymmetry study on a 24$\mu$m-selected sample ($L_{IR}
 \geq 5.0 \times 10^{10}\ L_{\odot}$), finding $m = 1.08$. However, these values are 
difficult to compare with ours because studies with selections in different bands yields 
different results (\citealt{bundy04,rawat08}; L09). 

\subsection{Merger fraction evolution at higher redshift}\label{ffhighz}
\begin{figure}[t]
\resizebox{\hsize}{!}{\includegraphics{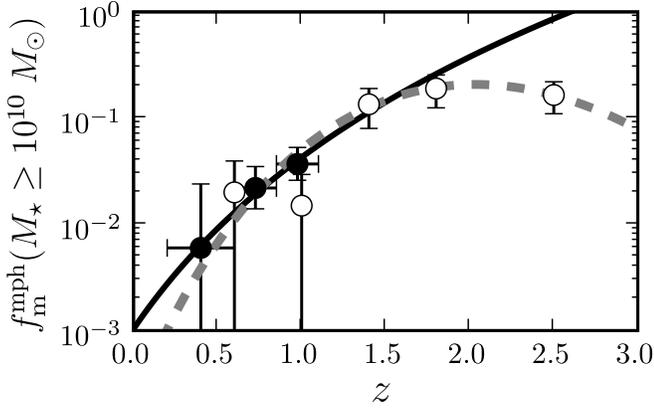}}
\caption{Morphological merger fraction vs redshift for $M_{\star}
 \geq 10^{10}\ M_{\odot}$ galaxies. Data are from \citet[][open circles]{conselice08}
 and this work (bullets). The black solid line gives the least-squares power-law fit 
to our data, $f_{\rm m}^{\rm mph}(z) = 0.001(1+z)^{5.4}$, while the dashed grey line 
is the least-squares fit to \citet{conselice08} data at $z > 1.2$ and ours at $z <
 1.2$, $f_{\rm m}^{\rm mph}(z) = 0.00034(1+z)^{10.5}{\rm e}^{-0.57(1+z)^2}$.\label{ffhighzfig}}
\label{ffhzfig}
\end{figure}

Merger fraction studies of $M_{\star} \geq 10^{10}\ M_{\odot}$ galaxies at 
redshift higher than $z \sim 1$ are rare. \citet{ryan08} address the problem with 
pair statistics, while \citet{conselice03ff,conselice08} use asymmetries. Both 
these studies conclude that the merger fraction shows a maximum at $z \gtrsim 1.5$ 
and decreases at higher $z$. This tells us that we cannot extrapolate the power-law 
fit (Eq.~[\ref{fffit}]) to high redshift. Fortunately, \citet{conselice08} perform 
their study by asymmetries, providing us with a suitably high redshift reference. Note 
that, although \citet{conselice08} treated the loss of information with redshift, they 
do not take into account the overestimation due to the experimental errors. Because the 
\citet{conselice08} study is performed in UDF, which is located in the GOODS-S area, we 
apply a 0.5 factor to the \citet{conselice08} merger fractions based on the results of
 Section~\ref{results}. In Fig.~\ref{ffhzfig} we show the corrected \citet{conselice08}
 merger fractions (white dots) and our data (black dots). Note that the 
previous power-law fit to our data (black solid line, Sect.~\ref{ffevolz}) fails 
to explain the merger fraction values at $z \gtrsim 1.5$.

Following \citet{conselice06ff}, we parameterize the observed tendency as
\begin{equation}
f_{\rm m}^{\rm mph}(z) = \alpha(1+z)^m{\rm e}^{\beta(1+z)^2}\label{fmbeta},
\end{equation}
where the local merger fraction is given by $f_{\rm m}^{\rm mph}(0) = 
\alpha\,{\rm exp}(\beta)$. This form is also obtained for the evolution 
of the merger rate on the basis of Press--Schechter theory \citep{carlberg90}. 
The best fit of the \citet{conselice08} data at $z > 1.2$ and ours at $z < 1.2$ 
yields $\alpha = 3.4\times10^{-4}$, $m = 10.5$, and $\beta = -0.57$ (gray
 dashed line in Fig.~\ref{ffhzfig}). With these values the merger fraction peaks 
at $z_{\rm peak} = 2$, in good agreement with \citet{conselice08}.

The previous parameterization implies that the merger fraction drops at
$z > z_{\rm peak} = 2$, being $\sim0.01$ at $z \sim 4$. On the other hand,
 \citet{hopkins08ff} models suggest that the merger fraction of $M_{\star} 
\geq 10^{10}\ M_{\odot}$ galaxies still grows at $z > z_{\rm peak}$, being 
$\sim0.30$ at $z \sim 4$. In fact, the data in Fig.~\ref{ffhzfig} can also been fitted by
\begin{equation}
f_{\rm m}^{\rm mph}(z) = \cases{0.001(1+z)^{5.4} & $z < z_{\rm c}$ \cr 
f_{\rm m,c}^{\rm mph} & $z \geq z_{\rm c}$}\label{fmc},
\end{equation}
where $f_{\rm m,c}^{\rm mph}$ is a constant, and $z_{\rm c}$ is the 
redshift in which the merger fraction behaviour changes; that is, when 
$0.001(1+z_{\rm c})^{5.4} = f_{\rm m,c}^{\rm mph}$. With the two high redshift 
points in Fig.~\ref{ffhighzfig} we estimate that $f_{\rm m,c}^{\rm mph} \sim 0.18 \pm 0.04$,
 which yields $z_{\rm c} = 1.62$. Further studies are needed to constrain the merger 
fraction evolution at high redshift, but it is clear that the potential approximation is only valid at $z \lesssim 1.5$.

\begin{figure}[t]
\resizebox{\hsize}{!}{\includegraphics{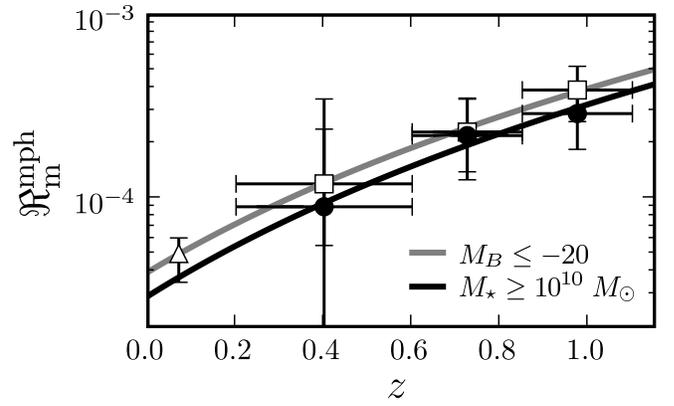}}
\caption{Morphological merger rate vs redshift for $M_B \leq -20$ (open 
triangle, \citealt{depropris07}; open squares, this work) and 
$M_{\star} > 10^{10}\ M_{\odot}$ galaxies (bullets). The grey/black solid 
line is the least-squares fit $\Re_{\rm m}^{\rm mph}(z) = \Re_{\rm m}^{\rm mph}(0) (1+z)^n$ 
to the data from the luminosity/mass-selected sample, respectively.}
\label{mrfig}
\end{figure}

\subsection{The major merger rate evolution}\label{mrgoods}
We define the major merger rate ($\Re_{\rm m}^{\rm mph}$) as the comoving 
number of major mergers per Gyr within a given redshift interval and luminosity 
or stellar mass range:
\begin{equation}
\Re_{\rm m}^{\rm mph} (z,M) = \rho(z,M) f_{\rm m}^{\rm mph}(z,M)T_{{\rm m},A}^{-1},
\end{equation}
where $M = M_B/M_{\star}$ denotes the selection of the sample, $\rho(z,M)$ 
is the comoving number density of galaxies at redshift $z$ brighter/more massive than 
$M_B/M_{\star}$, and $T_{{\rm m}, A}$ is the merger timescale in Gyr for the asymmetry 
criterion. To obtain $\rho(z,M_B)$ we assume the \citet{faber07} luminosity function 
parameters, while to obtain $n(z,M_{\star})$ we assume the \citet{pgon08} mass function 
parameters. In addition, we take $T_{{\rm m}, A} = 0.35 - 0.6$ Gyr. The lower value is
from \citet[][N-body major merger simulations]{conselice06ff}, and the higher from
 \citet[][N-body/hydrodynamical equal-mass merger simulations]{lotz08t}.

\begin{table*}
\caption{Major merger rates $\Re_{\rm m}^{\rm mph}$ in GOODS-S}
\label{mrtab}
\centering
\begin{tabular}{lccccc}
\hline\hline
Sample selection & $z = 0.4\ ^{\mathrm{a}}$ & $z = 0.725\ ^{\mathrm{a}}$ & 
$z = 0.975\ ^{\mathrm{a}}$ & $\Re_{\rm m}(0)\ ^{\mathrm{a,b}}$ & $n\ ^{\mathrm{b}}$\\
\hline
$M_B \leq -20$  &	$1.2^{+1.3}_{-0.6}$ & $2.3^{+1.4}_{-0.9}$ & $3.9^{+1.9}_{-1.3}$ &
 $0.40 \pm 0.14\ ^{\mathrm{c}}$ &  $3.3 \pm 0.8\ ^{\mathrm{c}}$\\
$M_{\star} \geq  10^{10}\ M_{\odot}$ & $0.9^{+2.6}_{-0.7}$ & $2.2^{+1.5}_{-0.9}$
 & $2.9^{+1.6}_{-1.0}$ & $0.29 \pm 0.06$ & $3.5 \pm 0.4$\\
\hline 
\end{tabular}
\begin{list}{}{}
\item[$^{\mathrm{a}}$]In units of $10^{-4}\ {\rm Mpc^{-3}\ Gyr^{-1}}$.
\item[$^{\mathrm{b}}$]Best $\Re_{\rm m}^{\rm mph}(z) = \Re_{\rm m}^{\rm mph}(0)(1+z)^n$ fit to the data.
\item[$^{\mathrm{c}}$]This fit includes the \cite{depropris07} local value.
\end{list}
\end{table*}

We summarize the merger rates in Table~\ref{mrtab} and show these values in 
Fig.~\ref{mrfig}: white symbols are for $M_B \leq -20$ galaxies (white 
triangle, \citealt{depropris07}; white squares, this work) and black 
dots for $M_{\star} \geq 10^{10}\ M_{\odot}$, while the grey/black solid line 
is the least-squares fit of $\Re_{\rm m}^{\rm mph} (z) = \Re_{\rm m}^{\rm mph}(0)(1+z)^n$ 
function to the data in the luminosity-/mass-selected sample. The parameters of these fits
 are also summarized in Table~\ref{mrtab}. In spite of the very different merger fraction 
evolution, the merger rate evolution of both samples are similar: $n = 3.3 \pm 0.8$ in the 
luminosity sample, while $n = 3.5 \pm 0.4$ in the mass sample. As in the merger fraction 
case, the results are not affected by the LSS (Sect.~\ref{lss}). The reason why the very
 different merger fraction evolution tuns into a similar merger rate evolution is the 
evolution over cosmic time of the number density of galaxies. The number of 
$M_B \leq -20$ galaxies {\it decreases} by a factor 3 from $z = 1$ to $z = 0,$ while
 the number of $M_{\star} \geq  10^{10}\ M_{\odot}$ galaxies {\it increases} by a factor 3
 in the same redshift range.

We can compare our inferred merger rate with the post-starburst (PSB) rate reported by \citet{wild09}. The light of PSB galaxies is dominated by A/F stars.  Such galaxies are identifiable by their strong Balmer absorption lines compared to their mean stellar age as measured by their 4000 \AA\ break strength. PSB spectra indicate that the formation of O- and early B-type stars has suddenly ceased in the galaxy. The simulations performed by \citet{johansson08} find that the PSB phase can only be reached by disc--disc major merger remnants, so the PSB rate and our merger rate may be similar if an evolutionary path connects both populations. The PSB rate, in the range $0.5 < z < 1$ and for $M_{\star} \gtrsim 10^{10}\ M_{\odot}$ galaxies (Salpeter IMF), is $\Re_{\rm PSB} =$ (1.6--2.9) $\times\ 10^{-4}\ {\rm Mpc^{-3}\ Gyr^{-1}}$, where the interval reflects the uncertainty in the PSB phase time-scale (0.35--0.6 Gyr, \citealt{wild09}). This value compares well with the inferred disc--disc major merger rate at that range, $\Re_{\rm m}^{\rm mph} = $ (1.2--3.0) $\times\ 10^{-4}\ {\rm Mpc^{-3}\ Gyr^{-1}}$. Although the uncertainties in both studies are important, the result suggests that SPB galaxies can be the descendants of our distorted, disc--disc major merger remnants.

\subsubsection{Number density of merger remnants}
If we integrate the merger rate over cosmic time, we obtain the number density 
of galaxies that have undergone a disc--disc major merger ($\rho_{\rm rem}$) in a given redshift range:
\begin{equation}\label{nrem}
\rho_{\rm rem}(z_1,z_2) = \int_{z_1}^{z_2}\Re_{\rm m}^{\rm mph}(0)(1+z)^{n-1}\frac{{\rm d}z}{H_0 E(z)},
\end{equation}
where $E(z) = \sqrt{\Omega_{\Lambda} + \Omega_{M}(1+z)^3}$ in a flat universe. 
We make this study only for the mass-selected sample because we can assume that stellar mass is additive:
$M_{\star}(z_1) \geq M_{\star}(z_2)$ always for $z_1 < z_2$, and $\rho_{\rm rem}(0,z)$ 
is representative of the number density of local galaxies that have undergone a
 disc--disc merger since redshift $z$. The same cannot be said for the luminosity-selected 
sample: here the number density of objects above a given absolute magnitude can decrease with
time, as it is not generally the case that $M_B(z_1) \leq M_B(z_2)$ for $z_1 < z_2$. Using Eq.~(\ref{nrem}) for $\rho_{\rm rem}(0,z)$ would overestimate the number of local galaxies that have undergone a merger. 

Comparing $\rho_{\rm rem}$ with the number of $M_{\star} \geq 10^{10}\ M_{\odot}$
 galaxies at redshift $z_1$, $\rho(z_1)$, we obtain the fraction of merger remnants,
\begin{equation}\label{frem}
f_{\rm rem}(z_1,z_2) = \frac{\rho_{\rm rem}(z_1,z_2)}{\rho(z_1)}.
\end{equation}
Applying Eq.~(\ref{frem}) with the merger rate parameters of the mass sample from
 Table~\ref{mrtab} and the mass functions from \citet{pgon08}, we obtain $f_{\rm rem}(0,1) =
 8^{+4}_{-3}$\%. This is a low value that increases to $f_{\rm rem}(0,1.5) = 15^{+9}_{-5}$\%.
 We take $z_2 = 1.5$ as an upper limit because our merger fraction parameterization is valid 
to this redshift (Sect.~\ref{ffhighz}) and \citet{pgon08} mass functions are complete for 
$M_{\star} \geq 10^{10}\ M_{\odot}$ galaxies also up to $z \sim 1.5$. Interestingly, we infer
 that $f_{\rm rem}(1.0,1.5) = 21^{+14}_{-9}$\%, which is compatible with the fraction of 
bulge-dominated galaxies (E/S0/Sa) at $z \sim 1$ (L\'opez-Sanjuan et al., in prep). The pair study of \citet{bundy09} in GOODS-S reports $f_{\rm rem}$ in the range 
$0.4 < z < 1.4$. For $M_{\star} \geq 2\times10^{10}\ M_{\odot}$ galaxies they estimate 
$f_{\rm rem}(0.4,1.4) = 15$\%-18\%, which is in good agreement with our inferred 
value, $f_{\rm rem}(0.4,1.4) \sim 17$\%. Given that the \citet{bundy09} study is also 
sensitive to mergers between spheroids, and the mass limit in both studies is different, the quantitative agreement is remarkable. 
 
The most important error source in our results is the uncertainty in the lower 
redshift bin, especially in the mass-selected sample. We repeat our study with a 
higher merger fraction in this bin by a factor of two, $f_{\rm m}^{\rm mph}(z = 0.4)
 = 0.012$, and three, $f_{\rm m}^{\rm mph}(z = 0.4) = 0.018$. With these assumptions 
$f_{\rm rem}(0,1)$ increases to 12\% and 18\%, respectively. These values remain low, 
so our conclusions do not change.

\subsubsection{Number of mergers per massive galaxy}
As a complement to the previous section we calculate the number of expected disc--disc 
major mergers per $M_{\star} \geq 10^{10}\ M_{\odot}$ galaxy in a given redshift range,
\begin{eqnarray}
N_{\rm m}(z_1,z_2) = \int_{z_1}^{z_2}\frac{\Re_{\rm m}^{\rm mph}(z)}{\rho(z)}\frac{{\rm d}z}{H_0 E(z)} \nonumber\\ 
= \int_{z_1}^{z_2}f_{\rm m}^{\rm mph}(0)(1+z)^{m-1}\frac{{\rm d}z}{T_{{\rm m},A} H_0 E(z)}.\label{nmerger}
\end{eqnarray}
Taking Eq.~(\ref{fmbeta}) as the merger fraction parameterization we obtain $N_{\rm m}(0,3) = 
1.2^{+0.4}_{-0.2}$. In addition, we also obtain $N_{\rm m}(1,3) = 1.0^{+0.4}_{-0.2}$, with 
only 0.2 disc--disc major mergers in $0 < z < 1$. The results are the same if we take 
Eq.~(\ref{fmc}) as the merger fraction parameterization. The previous $N_{\rm m}(0,3)$ 
value is lower than that inferred by \citet{bluck09}\footnote{We apply Eq.~(\ref{nmerger}) 
to their best power-law fit of the merger fraction and assume, as previously, that $T_{{\rm m},A} = 
0.35-0.6$ Gyr.} for $M_{\star} \geq 10^{11}\ M_{\odot}$ galaxies, $N_{\rm m}(0,3) = 1.8^{+0.6}_{-0.4}$, 
which implies that more massive galaxies have higher number of mergers than less massive ones. On the 
other hand, our value is $\sim 4$ times lower than that of \citet{conselice06ff,conselice08}, 
$N_{\rm gm}(0,3) \sim 4.4$ for $M_{\star} \geq 10^{10}\ M_{\odot}$ galaxies. We suspect that 
two factors contribute to their high value. First, they use the galaxy merger fraction 
($f_{\rm gm}$, the fraction of galaxies undergoing mergers) to obtain $N_{\rm gm}(z_1,z_2)$, 
the mean number of galaxies that merge since $z_2$ to obtain a $z_1$ galaxy. This is roughly $2$
 times higher than $N_{\rm m}$ (the number of merger events per galaxy). And second, they use 
classical counting statistics which, as shown in Sects~\ref{results} and \ref{ffhighz}, leads 
to an overestimate of the merger fraction by another factor of 2.

These results suggest that most of the disc--disc merger activity of 
$M_{\star} \geq 10^{10}\ M_{\odot}$ galaxies happened before $z \sim 1$, 
this kind of merger being important in galaxy evolution down to this redshift.
 It is important to recall that our methodology cannot detect spheroidal major 
mergers, so the role of these mergers in the evolution of the red sequence since 
$z \sim 1$ \citep{bell04,faber07} cannot be addressed by our study. However,
 due to the paucity of spheroidal systems at $z \gtrsim 1.2$ \citep{conselice05,cassata05},
 one expects spheroidal major mergers to be important at lower redshifts; 
i.e.\ $z \lesssim 1.2$. The simulations of \citet{khochfar08} are in agreement with
 this picture: they find that the dry merger rate is two orders of magnitude less than the 
wet merger rate at $z \sim 1.5$, while they are similar at $z \sim 0.6$. In addition, they 
find that the wet merger rate has its maximum at $z \sim 1.3$, and then declines by an order
 of magnitude until $z = 0$, a similar evolution to our results, 
$\Re_{\rm m}^{\rm mph} (1.3) \sim 20\Re_{\rm m}^{\rm mph} (0)$. To check these 
ideas we explore the relative importance of disc--disc mergers in the structural 
evolution of $M_{\star} \geq 10^{10}\ M_{\odot}$ galaxies in a forthcoming paper.

\begin{figure}[t]
\resizebox{\hsize}{!}{\includegraphics{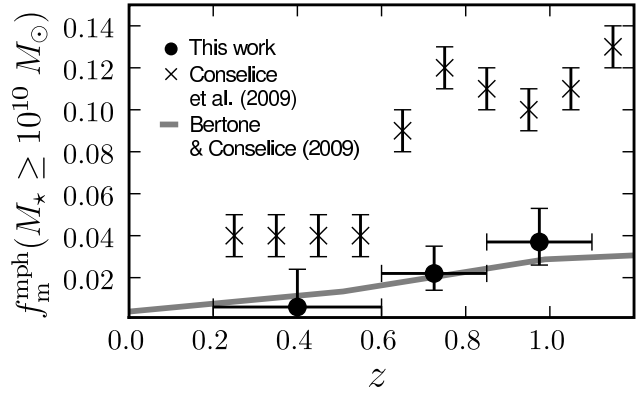}}
\caption{Comparison between the observed and simulated merger fractions for galaxies with $M_{\star} \geq 10^{10}\ M_{\odot}$. The observations are from this work (black bullets), and from \citet[][crosses]{conselice09cos}. The predictions (gray solid lines) are for major mergers with a time-scale of $T_{{\rm m}, A} = 0.4$ Gyr \citep{bertone09}.}
\label{fmssim}
\end{figure}

\subsection{Comparison with model predictions}
The comparison of the predicions by cosmological simulations with our results is not straighforward because we only detect disk-disk (i.e.\ wet) major mergers, and we  select by stellar mass: the simulations from \citet{stewart08} point out
 that the merger fraction depends on merger definition (minor vs major), selection criteria (halo mass, stellar mass, or luminosity) or the assumed merger time-scale.

The study from \citet{bertone09} provides predictions for major mergers of $M_{\star} \geq 10^{10}\ M_{\odot}$ galaxies, assuming a merger time-scale of $T_{{\rm m}, A} = 0.4$ Gyr. In figure~\ref{fmssim} we show the predictions (gray solid line), and the observational data from this work (black bullets) and \citet[][crosses]{conselice09cos}. The predictions are in good agreement with our observations, while the \citet{conselice09cos} values are higher than predicted by factors 2 to 6. However, this agreement must be taken as qualitative more than quantitative because (i) cosmological simulations might underestimate the major merger fraction at that stellar mass, as pointed out by \citet{bertone09}, (ii) the predictions are for {\it total} (i.e.\ wet + dry) major mergers, while we report wet major mergers. This can lead in a $\sim$1\% increase in the merger fractions due to dry mergers \citep{bell06ee,lotz08t}. And, (iii) the GOODS-S merger fractions might be lower that the cosmological value due to cosmic variance (Sect.~\ref{cosvar}). Despite these caveats, the agreement is remarkable.

On the other hand, the simulations of \citet{stewart09} provide $f_{\rm rem}(0,2)$ for major wet mergers in $M_{\star} \geq 10^{10}\ M_{\odot}$ galaxies: they predict $f_{\rm rem}(0,2) \sim 10$\%-20\%, in good agreement with our inferred $f_{\rm rem}(0,2) \sim 21^{+11}_{-7}$\%. Finally, \cite{wein09} compare their study of $M_{\star} \geq 10^{10}\ M_{\odot}$ local spiral galaxies  with the predictions by the \citet{khochfar06} and \citet{hopkins09bulges} models. They find that both models are able to explain the observed bulge-to-total ratio ($B/T$) distribution, and predict that only 13--16\% of today's $B/T < 0.75$ spirals have undergone a major merger since $z = 2$. If we assume that all the disc--disc major mergers since $z = 2$ have enough gas to re-form a disc in the merger remnant \citep{hopkins09disk}, our $f_{\rm rem}(0,2) \sim 20$\% value is an upper limit to the models' predictions, so both are compatible with our results.
 
\section{CONCLUSIONS}\label{conclusion}
We have computed the disc--disc major merger fraction and its evolution up to 
$z \sim 1$ in the GOODS-S field using morphological criteria. We quantify and 
correct for the bias due to varying spatial resolution and image depth with redshift
 by artificially redshifting the galaxy images to a common reference redshift of $z_{\rm d} = 1$.
 More importantly, we successfully account for the spill-over of sources into
 neighbouring bins caused by the errors in asymmetry indices and in $z_{\rm phot}$,
 through the use of an ML method developed by LGB08. In every case we obtain merger
 fractions lower than $0.06$, in agreement with the merger fraction determination for 
the Groth field (L09). The main improvement 
in our study over previous determinations is the robust methodology that takes into 
account the signal-to-noise variation of galaxies with $z$ and the observational errors: 
previous morphological studies using classical counts overestimate the disc--disc major merger fractions by factors of $\sim 2$.

The merger fraction evolution in luminosity- and mass-selected
 samples are, respectively,
\begin{equation}
f_{\rm m}^{\rm mph}(z, M_B \leq -20) = 0.013(1+z)^{1.8},
\end{equation}
\begin{equation}
f_{\rm m}^{\rm mph}(z, M_{\star} \geq 10^{10}\ M_{\odot}) = 0.001(1+z)^{5.4}.
\end{equation}
We study the effect of the LSS on these results and find that merger fractions do not change substantially.

When we compute the merger rate for both samples, the very different merger fraction
 evolution becomes a quite similar merger rate evolution:
\begin{equation}
\Re_{\rm m}^{\rm mph}(z) = 0.40\times10^{-4}(1+z)^{3.3}\ {\rm Mpc^{-3}Gyr^{-1}}
\end{equation}
for $M_B \leq -20$ galaxies and
\begin{equation}
\Re_{\rm m}^{\rm mph}(z) = 0.29\times10^{-4}(1+z)^{3.5}\ {\rm Mpc^{-3}Gyr^{-1}}
\end{equation}
for $M_{\star} \geq 10^{10}\ M_{\odot}$ galaxies. This similar evolution is due to 
the different number density evolution with redshift: the number of 
$M_{\star} \geq 10^{10}\ M_{\odot}$ galaxies increases with cosmic time, while the
 number of $M_B\leq -20$ galaxies decreases.

The previous merger rates imply that only $\sim 8$\% of today's $M_{\star} \geq 10^{10}\ M_{\odot}$
 galaxies have undergone a disc--disc major merger since $z\sim1$. Interestingly, $\sim 21$\% of 
these galaxies  at $z \sim 1$ have undergone a disc--disc major merger since $z\sim1.5$, which 
is compatible with the fraction of bulge-dominated galaxies (E/S0/Sa) at $z \sim 1$ 
(L\'opez-Sanjuan et al., in prep). This suggests that disc--disc major mergers are not 
the dominant process in evolution of $M_{\star} \geq 10^{10}\ M_{\odot}$ galaxies since 
$z\sim1$, with only 0.2 disc--disc major mergers per galaxy, but may be an important 
process at $z > 1$, with $\sim 1$ merger per galaxy at $1 < z < 3$.

The most important error source in these results is the uncertainty in the lower 
redshift bin, especially in the mass-selected sample. More studies are needed to improve the statistics at low redshift 
and avoid cosmic variance effects. Another important issue is the sample definition, given that merger fraction depends on mass and luminosity: larger samples permit us 
different selection cuts in luminosity and mass, thus improving our knowledge of the 
importance of disc--disc major mergers in galaxy evolution.

\begin{acknowledgements}
We dedicate this paper to the memory of our six IAC colleagues and friends who 
met with a fatal accident in Piedra de los Cochinos, Tenerife, in February 2007,
 with particular thanks to Maurizio Panniello, whose teaching of \texttt{python}
 was so important for this paper.

We thank the anonymous referee for suggestions that improved the paper. This work was supported by the Spanish Programa Nacional de Astronom\'\i a y 
Astrof\'{\i}sica through project number AYA2006--12955, AYA2006--02358 and
AYA 2006--15698--C02--02. This work was partially funded by the Spanish MEC under 
the Consolider-Ingenio 2010 Program grant CSD2006-00070: First Science with the GTC 
(http://www.iac.es/consolider-ingenio-gtc/). This work is based on {\it HST/ACS} images from GOODS {\it HST} Treasury Program, which is supported by NASA throught grants HST-GO-09425.01-A and HST-GO-09583.01, and in part on observations made with the {\it Spitzer} Space Telescope, which is operated by the Jet Propulsion Laboratory, Caltech under NASA contract 1407. 

P. G. P. G. acknowledges support from the Ram\'on y Cajal Program financed by the Spanish Government and the European Union.

\end{acknowledgements}

\bibliography{aamnemonic,1923bib}
\bibliographystyle{aa}
\end{document}